\documentclass[fleqn,usenatbib]{mnras}

\usepackage{newtxtext,newtxmath}

\usepackage[T1]{fontenc}
\usepackage{ae,aecompl}

\usepackage{graphicx}	
\usepackage{amsmath}	
\usepackage{siunitx}
\usepackage{multicol}
\usepackage{enumitem}

\newcommand{\equ}[1]{eq.~(\ref{eq:#1})}
\newcommand{\equs}[1]{eqs.~(\ref{eq:#1})}
\newcommand{\equm}[1]{(\ref{eq:#1})}

\newcommand{\equnp}[1]{eq.~\ref{eq:#1}}
\newcommand{\equsnp}[1]{eqs.~\ref{eq:#1}}
\newcommand{\equmnp}[1]{\ref{eq:#1}}
\newcommand{\se}[1]{Section~\ref{sec:#1}}
\newcommand{\fig}[1]{Figure~\ref{fig:#1}}

\newcommand{\tab}[1]{Table~\ref{tab:#1}}
\newcommand{\be}{\begin{equation}}
\newcommand{\ee}{\end{equation}}
\newcommand{\bea}{\begin{eqnarray}}
\newcommand{\eea}{\end{eqnarray}}

\newcommand{\no}{\noindent}

\newcommand{\msun}{{\rm M}_\odot}
\newcommand{\Msun}{M_\odot}

\newcommand{\ifm}[1]{\relax\ifmmode#1\else$\mathsurround=0pt #1$\fi}
\newcommand{\kms}{\ifmmode\,{\rm km}\,{\rm s}^{-1}\else km$\,$s$^{-1}$\fi}

\newcommand{\kpc}{\,{\rm kpc}}
\newcommand{\pc}{\,{\rm pc}}

\newcommand{\K}{\,{\rm K}}
\newcommand{\cmc}{\,{\rm cm^{-3}}}

\newcommand{\ltsima}{$\; \buildrel < \over \sim \;$}
\newcommand{\lsim}{\lower.5ex\hbox{\ltsima}}
\newcommand{\gtsima}{$\; \buildrel > \over \sim \;$}
\newcommand{\gsim}{\lower.5ex\hbox{\gtsima}}

\def\la{\langle}

\def\cmc{\,{\rm cm}^{-3}}

\def\M*{M_{\rm *}}
\def\Mv{M_{\rm vir}}
\def\Rv{R_{\rm vir}}
\def\Vv{V_{\rm vir}}
\def\Vs{V_{\rm s}}

\def\tcm{t_{\rm cool,\,mix}}
\def\Tv{T_{\rm vir}}
\def\Ts{T_{\rm s}}
\def\Tb{T_{\rm b}}
\def\Rs{R_{\rm s}}
\def\rhob{\rho_{\rm b}}
\def\rhos{\rho_{\rm s}}

\def\cb{c_{\rm b}}
\def\cs{c_{\rm s}}

\def\tsc{t_{\rm sc}}
\def\Mb{M_{\rm b}}

\def\Pi{\varpi_{_{\rm I}}}

\title[OVI Traces Photoionized Streams With Collisionally Ionized Boundaries in Simulations at $z\sim1$]{OVI Traces Photoionized Streams With Collisionally Ionized Boundaries in Cosmological Simulations of $z\sim1$ Massive Galaxies}
\author[C. Strawn et al.]{Clayton Strawn$^{1}$\thanks{E-mail: cjstrawn@ucsc.edu}, Santi Roca-F\`abrega$^{2}$, Nir Mandelker$^{3,4,5,6}$, Joel Primack$^{1}$, \newauthor Jonathan Stern$^{7}$, Daniel Ceverino$^{8}$, Avishai Dekel$^{6,1}$, Bryan Wang$^{9}$, \newauthor Rishi Dange$^{9}$\\
\smallskip 
$^1$ University of California, Santa Cruz, CA 95064, USA\\
$^2$ Universidad Complutense de Madrid, Departamento de F\'isica de la Tierra y Astrof\'isica, Madrid 28040, Spain\\ 
$^3$ Department of Astronomy, Yale University, P.O. Box 208101, New Haven, CT, USA\\
$^4$ Heidelberger Institut f\"ur Theoretische Studien, Schloss-Wolfsbrunnenweg 35, D-69118 Heidelberg, Germany\\
$^5$ Kavli Institute for Theoretical Physics, University of California, Santa Barbara, CA 93106, USA\\
$^6$ Racah Institute of Physics, The Hebrew University of Jerusalem, Jerusalem 91904, Israel\\
$^7$ Northwestern University, IL 60208, USA\\
$^8$ CIAFF, Facultad de Ciencias, Universidad Aut\'onoma de Madrid, Madrid 28049, Spain\\
$^9$ The Harker School, CA 95129, USA}
\date{Accepted XXX. Received YYY; in original form ZZZ}

\pubyear{2020}

\begin{document}

\defcitealias{stern_universal_2016}{S16}
\defcitealias{stern_does_2018}{S18}
\defcitealias{roca-fabrega_cgm_2019}{RF19}
\defcitealias{mandelker_instability_2020}{M20a}
\defcitealias{mandelker_lyalpha_2020}{M20b}

\label{firstpage}
\pagerange{\pageref{firstpage}--\pageref{lastpage}}
\maketitle

\begin{abstract}
\smallskip 
We analyse the distribution and origin of OVI in the Circumgalactic Medium (CGM) of dark-matter haloes of $\sim 10^{12}$ M$_\odot$ at $z\sim1$ in the VELA cosmological zoom-in simulations. We find that the OVI in the inflowing cold streams is primarily photoionized, while in the bulk volume it is primarily collisionally ionized. The photoionized component dominates the observed column density at large impact parameters ($\gtrsim 0.3 R_{\rm vir}$), while the collisionally ionized component dominates closer in. We find that most of the collisional OVI, by mass, resides in the relatively thin boundaries of the photoionized streams. Thus, we predict that a reason previous work has found the ionization mechanism of OVI so difficult to determine is because the distinction between the two methods coincides with the distinction between two significant phases of the CGM.  We discuss how the results are in agreement with analytic predictions of stream and boundary properties, and their compatibility with observations. This allows us to predict the profiles of OVI and other ions in future CGM observations and provides a toy model for interpreting them.
\end{abstract}




\section{Introduction}

The analysis of the Circumgalactic Medium (CGM), the gas that resides outside galactic discs but still within or near its virial radius, has the potential of giving us valuable information about the past history of galaxy formation and also some hints of its future evolution \citep[e.g. ][]{tumlinson_circumgalactic_2017}. The importance of studying the CGM in order to characterize the history of galaxy formation is clear: it has been shown conclusively that galaxies alone contain significantly fewer baryons than would be expected from the standard $\Lambda$CDM cosmology (the `missing baryon problem'). While a significant fraction of these baryons may have been ejected to the Intergalactic Medium (IGM) in the early stages of galaxy formation \citep{Aguirre2001}, or remained as warm-hot low-metallicity intergalactic gas throughout cosmic time \citep{Shull2012}, studies suggest that 10-100 percent of the cosmic baryon budget of the universe exists in the metal-rich CGM of galactic halos \citep{werk_cos-halos_2014,bordoloi_cos-dwarfs_2014}. The CGM is thus certainly significant and possibly even dominant in baryonic matter. Further, by calculations of the total amount of metals produced in all stars, only about 20-25 percent remain in the stars, ISM gas, and dust \citep{peeples_budget_2014}. Recent studies have been consistent with the idea that most of metals produced within stars and released by supernovae feedback or stellar winds reside in the metal-rich CGM \citep{tumlinson_large_2011,werk_cos-halos_2013}. The mechanisms collectively known as `feedback' by which metals, mass, and energy are transported to the CGM, are not yet completely understood, and likely include contributions from several processes like stellar winds, supernovae feedback, and interaction with winds from the central AGN. The interactions between these feedback mechanisms and their relative contributions, and their dependence on halo mass and redshift might be better constrained by studies of the kinematics and temperatures of the ions within the CGM. 

The CGM properties are also relevant when studying the future evolution of galaxies. Current models of the $z \gtrsim 1$ CGM state that cold, relatively low metallicity gas inflows from the IGM feed star formation of central galaxies through narrow streams \citep{keres_how_2005,dekel_galaxy_2006,dekel_cold_2009,ocvirk_bimodal_2008}. An overview of the topic is found in \citet{fox_gas_2017}, and references therein. Outside those streams, metal-enriched warm-hot gas (10$^{4.5}$~K$<$T$<$10$^{6.5}$~K) mainly produced by stellar feedback from the central galaxy, and by the virial shock under specific conditions \citep{white_core_1978,birnboim_virial_2003,fielding_impact_2017,stern_maximum_2020}, fills the rest of the CGM volume. Although widely accepted by the community, these models still suffer from large uncertainties due to the difficulty of observations and of comparing them with numerical simulations or analytic models. A more detailed review of this theoretical picture will also be presented in Section \ref{sec:streams_theory}.

A useful parameter for the interpretation of existing and future data for those larger-scale phenomena is the ionization state of gas within the CGM, which as a rule is highly ionized \citep{werk_cos-halos_2014}. While the ionization level of the gas within the CGM can help to constrain the physical interpretation, analyzing the full volume of gas from a single or even several ion species is remarkably difficult \citep{tumlinson_circumgalactic_2017}. This is because atoms can be ionized, in general, in two different ways. They can be photoionized (PI), meaning incoming photons from either the galaxy itself or the ultraviolet background light interact with an atom and strip it of electrons, or they can be collisionally ionized (CI), meaning thermalized interactions with nearby atoms will `knock off' electrons, leaving the atoms in some distribution of ionization states \citep{osterbrock_astrophysics_2006}. Broadly speaking, PI gas fractions are a function of density (as denser gas recombines more quickly, biasing gas towards lower ionization states) and CI gas fractions are a function of temperature (as hotter gas will have more kinetic energy per particle, biasing gas towards higher ionization states). Most studies tend to assume that only one of these mechanisms is in play at a time for a given patch of gas, with the rationalization that since denser gas tends also to be cooler, either mechanism will result in cold gas hosting low-ions (e.g. NII, MgII) and hot gas hosting high ions (e.g. NeVIII, MgX). Examples of assuming OVI is in PI-equilibrium include \citet{stern_universal_2016} and assuming OVI is in CI-equilibrium include \citet{faerman_massive_2017}. However, such an assumption is very tenuous. In \citet{roca-fabrega_cgm_2019} (hereafter \citetalias{roca-fabrega_cgm_2019}), it was found that in cosmological simulations of halos which reached roughly Milky-Way masses at $z \sim 1$, whether OVI is PI or CI depends strongly on redshift, mass, and position within the CGM, and ranged all the way from $\sim$ 100 percent PI to $\sim$ 0 percent PI over the course of their evolution. The main conclusions of \citetalias{roca-fabrega_cgm_2019} were that OVI is more photoionized in the outer halo than the inner, and that galaxies transform from fully collisionally ionized to mostly photoionized at z=2, after which they diverge by mass, with larger galaxies becoming more collisionally ionized than smaller ones.

Detections of very high ionization states such as OVII and OVIII in the MW \citep{gupta_huge_2012,fang_xmm-newton_2015} indicate that there is indeed a significant warm-hot component of the CGM, and it is a source of major controversy whether OVI should be considered to mostly be cospatial with that gas, whether it should be considered to be mostly cool and more closely connected to H I and low metal ion states, or whether both are relevant simultaneously. We will especially consider \citet{stern_universal_2016} and \citet{stern_does_2018} (hereafter \citetalias{stern_universal_2016}
 and \citetalias{stern_does_2018}, respectively). In \citetalias{stern_universal_2016} a phenomenological model is proposed which explains the relations between low, intermediate, and high ionization states as a consequence of hierarchical PI densities, where smaller, denser spherical clouds containing low ions are embedded within larger, less dense clouds containing OVI. This model matched the observed absorption much better than an assumption of a single or small number of densities, and nearly as well as models with a separate gas phase for each ion, with many fewer parameters (see \citetalias{stern_universal_2016}, Figure 6). \citetalias{stern_does_2018}, focusing especially on OVI, assumed this hierarchical density strcuture is global, with OVI residing in the outer halo and the low ions residing in the inner halo. They claimed that the majority of the OVI gas detected is located outside $\sim0.6R_{\rm vir}$. This radial distance, which is defined as the approximate radius of the median OVI particle, is called $R_{\rm OVI}$. In \citetalias{stern_does_2018}, both a `high-pressure' (CI) and a `low-pressure' (PI) scenario are presented which are consistent with the data. However, the PI scenario more naturally explains the observed $N_{\rm HI}/N_{\rm OVI}$ values of $1-3$ seen at large impact parameters, and also alleviates the large energy input in the outer halo required by the CI scenario (see also \citealp{mathews_circumgalactic_2017} and \citealp{mcquinn_implications_2018}). It has also been suggested that the CGM might have both phases present at the same time in different regions, most recently in \citet{wu_constraints_2020}.

Due to the low emission measure of the CGM, it is necessary to do most observations through absorption spectra of bright background objects which have lines of sight passing through the CGM of intervening galaxies. While many studies made great strides towards understanding the CGM through absorption \citep[e.g.][]{tripp_missing_2004,rupke_outflows_2005,danforth_low-z_2005}, many of these studies had either a low number of available sources or had difficulty associating each absorber with an individual galaxy. In recent years with the increased sensitivity from the Cosmic Origins Spectrograph (COS) on Hubble Space Telescope \citep{werk_cos-halos_2013,werk_cos-halos_2014,tumlinson_large_2011,prochaska_quasars_2013,bordoloi_cos-dwarfs_2014,burchett_deep_2016}, there has been an explosion of absorption line studies. This analysis can be used to detect relatively small column densities for different ions, including metal lines, giving insight into the temperature, density, and metallicity information of the gas. However, there are important limitations to this type of analysis. First and foremost, absorption-line studies require a relatively rare alignment between the background source and the foreground galaxy, and there are only a few examples where several lines pass through the same CGM in different places \citep{lehner_evidence_2015,bowen_structure_2016} or strong gravitational lensing allows the same source to be seen in multiple locations throughout the CGM \citep{lopez_clumpy_2018,okoshi_multiple_2019}. This means many assumptions must be made in order to combine together the data even within an individual survey, such as assuming relatively similar conditions of the CGM in all L$_*$ galaxies, and that the CGM is spherically symmetric (or at least isotropic). The use and results of those assumptions are examined in (\citealp{werk_cos-halos_2014,mathews_circumgalactic_2017}; \citetalias{stern_universal_2016}). Finally, the lack of visual imaging makes it very difficult to constrain the position of the detected gas along the line of sight, and it may well be at any distance greater than the impact parameter, or gas from different ions may even be in completely separate clouds. 

Numerical simulations are thus playing an important role as tools for testing recent theoretical approaches that try to characterize the CGM properties and evolution \citep[e.g.][]{shen_circumgalactic_2013,faerman_massive_2017,stern_does_2018,nelson_abundance_2018,roca-fabrega_cgm_2019,stern_cooling_2019,stern_maximum_2020}. Hydrodynamic simulations are commonly used to supplement analytical models regarding the CGM, as they can break the degeneracy between CI and PI gas. While most cosmological simulations have difficulty resolving the CGM due to the Lagrangian nature of the adaptive resolution, where the spatial resolution becomes very poor in the low density CGM/IGM \citep[e.g.][]{nelson_zooming_2016}, several recent groups have implemented novel methods to significantly enhance the resolution in the CGM, obtaining results similar to COS-Halos  and other observations \citep{hummels_impact_2019,peeples_figuring_2019,suresh_zooming_2019,van_de_voort_cosmological_2019,mandelker_shattering_2019,corlies_figuring_2020}. In a broad sense, the CGM remains a useful testing ground for these simulations, as the ionization state of the gas, as well as its phase (e.g. cold or hot mode accretion) will depend sensitively on the feedback mechanisms incorporated into the simulation, as well as the refinement algorithm and implementation of the code itself (\citealp{nelson_moving_2013}; see also \citealp{kim_agora_2013}, \citeyear{kim_agora_2016}). Several other hydrodynamic simulations have begun analyzing the OVI population in the CGM as well, including \citet{suresh_ovi_2017} and \citet{corlies_empirically_2016}, generally finding that this ion exists in a multiphase medium, and replicating the bimodality between high column densities in star-forming galaxies and low column densities in quenched galaxies, first seen in \citet{tumlinson_large_2011}.

In this work we analyse galaxies from the VELA simulation suite \citep{ceverino_radiative_2014,zolotov_compaction_2015}. These simulations are compared with observations in a number of papers \citep[e.g.][]{Ceverino15,Tacchella16b,Tacchella16a,Tomassetti16,mandelker_giant_2017,Huertas-Company19,Dekel20a,Dekel20b}, showing that many features of galaxy evolution traced by these simulations agree with observations, although the VELA simulated galaxies form stars somewhat earlier than observed galaxies. Using these simulations, we develop a model in which CGM gas is characterized by its relation to the aforementioned `cold accretion streams', giving OVI a unique role where PI OVI gas acts as an indicator of these inflows. Much of the CI OVI gas turns out to lie in an interface layer between these inflowing cold streams and the bulk of the CGM. 

This paper is organized as follows. In Section \ref{sec:simulation} we describe the simulation suite and our analysis methods. In Section \ref{sec:definitions} we explain how we robustly differentiate CI and PI gas for a variety of ions. Focusing on OVI, we analyse in Section \ref{sec:sec4overall} what this distinction shows us about the CGM and its structure. This is analyzed in 3D space in Section \ref{sec:3Ddistribution}, in projections and sightlines in Section \ref{sec:sightlines}, and its dependence on redshift and galaxy mass in Section \ref{sec:mass-redshift-dependence}. In Section \ref{sec:observations} we compare the implications of this model to the findings in \citetalias{stern_universal_2016} and \citetalias{stern_does_2018}. In Section \ref{sec:model_2} we present a physical model for the origin and properties of the different OVI phases in the CGM, relating these to cold streams interacting with the hot CGM. This is the first discussion of shear layer width around radiatively-cooling cold streams in galaxy halos. In Section \ref{sec:streams_theory} we summarize our current theoretical understanding of the evolution of cold streams in the CGM of massive high-z galaxies, as the streams interact with the ambient hot gaseous halo. In Section \ref{sec:streams_sims} we examine the properties of the different CGM phases identified in our simulations, in light of this theoretical framework. Finally, in Section \ref{sec:other_ions} we use these insights to model the distribution of OVI and other ions in the CGM of massive $z \sim 1$ galaxies. Our summary and conclusions are presented in Section \ref{sec:conclusion}.

\section{Data and analysis tools}\label{sec:simulation}
\subsection{VELA}
\begin{figure}
   \centering
      \includegraphics[trim={1.2cm 0.6cm 2.5cm 2.0cm}, clip, width=0.99\linewidth]{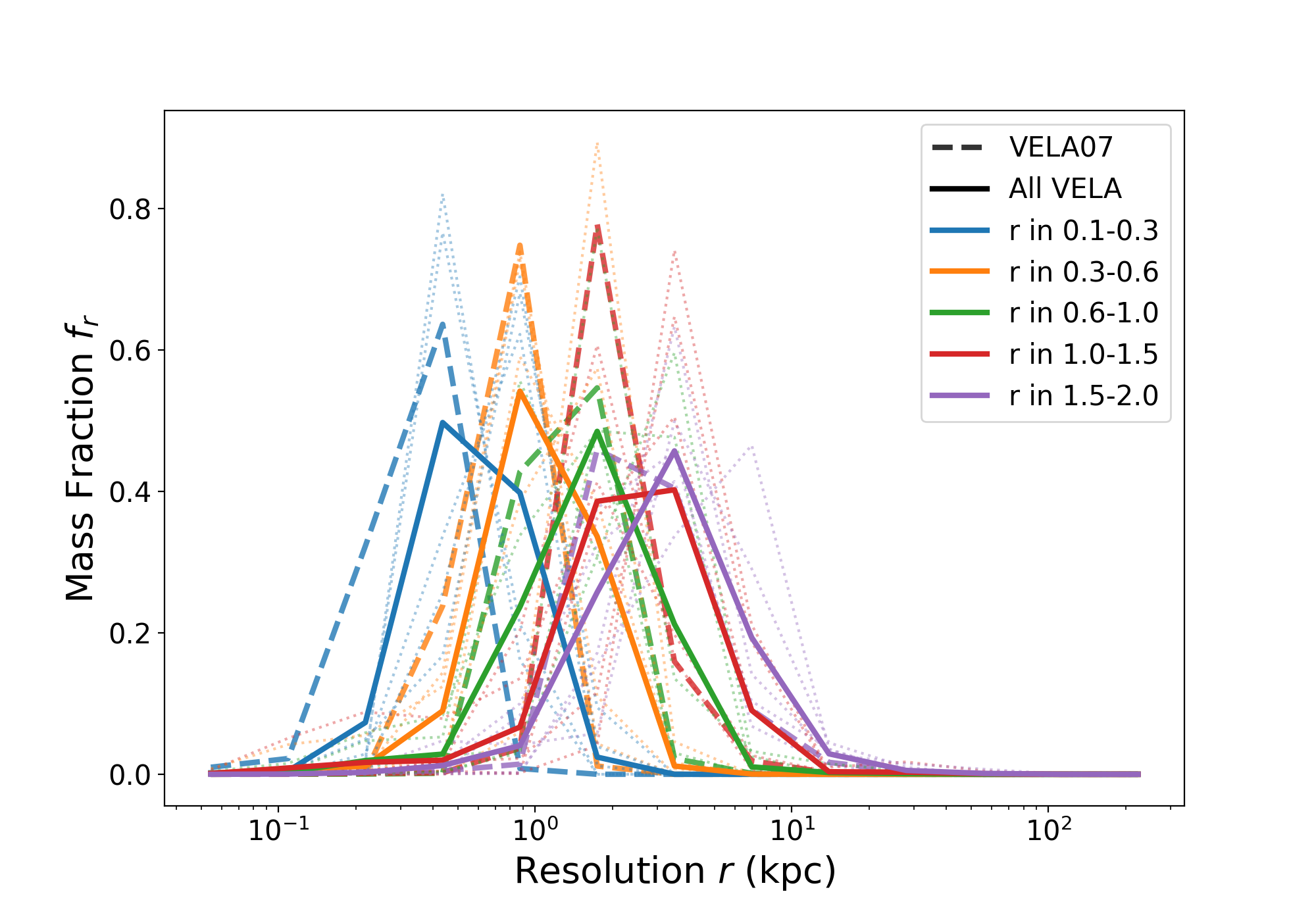}
   \caption{The mass fraction of gas in cubical cells of a specific side length. The different colors represent different distances from the galaxy center, in units of Rvir. Solid lines are an average over all 6 analysed VELA galaxies, dashed lines are from VELA07, for comparison with Figures \ref{fig:OVIslice} and  \ref{fig:ci-frac}. Semi-transparent lines represent individual results from the other selected VELA galaxies.}
   \label{fig:resolution}
\end{figure}

The set of VELA simulations we used is a subsample of 6 galaxies from the full VELA suite (see Table \ref{tab:1} for details about the galaxies chosen). The entire VELA suite contains 35 haloes with virial masses (M$_{\textrm v}$)\footnote{All virial quantities we show in this paper are taken from \citet{mandelker_giant_2017} and correspondence with the authors. They are calculated according to the definition of virial radius from \citet{bryan_statistical_1998}. This gives some disagreement between the numbers listed here and in \citetalias{roca-fabrega_cgm_2019}, where the $R_{200}$ value was used instead. $M_*$ only includes star particles within 10 kpc in order to ignore satellite galaxies within the halo.} between 2$\times$10$^{11}$M$_{\odot}$ and 2$\times$10$^{12}$M$_{\odot}$ at $z=1$. The VELA suite was created using the ART code \citep{kravtsov_adaptive_1997,kravtsov_origin_2003,ceverino_role_2009}, which uses an adaptive mesh with best resolution between 17 and 35 physical pc at all times. In the CGM, the resolution is significantly worse than this maximum, as expected. However, most of the mass within the virial radius is actually found to be in cells of resolution better than 2 kpc, as shown in Figure \ref{fig:resolution}. This is within an order of magnitude of several high-resolution CGM simulations of recent years \citep{peeples_figuring_2019,suresh_zooming_2019,hummels_impact_2019,van_de_voort_cosmological_2019,bennett_resolving_2020}, although unlike VELA, those simulations required these high resolutions throughout the CGM. This gives the VELA simulations enough resolution for discussions of the CGM to be physically meaningful, at least with respect to higher ions such as OVI which should be less dependent on resolution effects than low ions which likely originate from small clouds (\citealt{hummels_impact_2019}, see also \citetalias{stern_universal_2016}). 
Alongside gravity and hydrodynamics processes, subgrid models incorporate metal and molecular cooling, star formation, and supernova feedback \citep{ceverino_role_2009,ceverino_high-redshift_2010,ceverino_radiative_2014}. Star formation occurs only in cold, dense gas ($n_{\rm H}>1~$cm$^{-3}$ and T$<$10$^4$~K). In addition to thermal-energy supernova feedback the simulations incorporate radiative feedback from stars, adding a non-thermal radiation pressure to the total gas pressure in regions where ionizing photons from massive stars are produced. Recently, the VELA simulations have been re-run with increased supernova feedback \citep[following][]{gentry_enhanced_2017}, and this new feedback mechanism has led to improved stellar mass-halo mass relations as in Ceverino et al. (2020 in prep). We will compare the results of this paper to this newer version in future work. In the VELA simulations, the dark matter particles have masses of $8.3\times10^4$M$_{\odot}$, while the average star particle has a mass of $10^3$ M$_{\odot}$. Further details about the VELA suite can be found in \citep{ceverino_radiative_2014,zolotov_compaction_2015}.  

We chose to continue to use the same subsample of the VELA galaxies from \citetalias{roca-fabrega_cgm_2019}. In that work, they were chosen according to their virial masses and the final redshift the simulation reached. This means, we use all halos that have been simulated down to z = 1 which have a final mass greater than $10^{11.5} $M$_\odot$. This selection criteria derived from our desire to analyse the physical state of gas in galaxies near the `critical mass' at which the volume-filling CGM phases show a transition from free-fall to pressure-support \citep{birnboim_virial_2003,goerdt_distribution_2015,zolotov_compaction_2015,fielding_impact_2017,stern_virialization_2020}.

\begin{table}
\begin{centering}
 \begin{tabular}{l c c c c} 
 \hline\hline
VELA  & $M_{\rm vir}$ & $M_*$ & $M_{\rm gas}$ & $R_{\rm vir}$ \\
  & [$10^{12}$M$_{\odot}$] & [$10^{10}$M$_{\odot}$] & [$10^{10}$M$_{\odot}$] & [kpc] \\
 \hline
 {\bf V07} & {\bf 1.51} & {\bf 10.1} & {\bf 6.88} & {\bf 183} \\ 
 V08 & 0.72 & 2.93 & 4.80 & 132 \\
 V10 & 0.73 & 2.44 & 4.82 & 142 \\
 V21 & 0.86 & 6.54 & 3.09 & 152 \\
 V22 & 0.62 & 4.64 & 1.27 & 136 \\
 V29 & 0.90 & 3.73 & 5.03 & 152 \\
 \hline\hline
 \end{tabular}
  \caption{Properties at $z=1$ of the VELA simulations used in this work. This work uses the same set of galaxies as \citetalias{roca-fabrega_cgm_2019}. $M_*$ is defined as all stars within 10 kpc.}
\label{tab:1}
 \end{centering}
\end{table}
        
\subsection{Analytical approach and analysis tools}\label{sec:approach}

In our analysis of the CGM ionization state, we will study both the photoionization and the collisional ionization mechanisms. We will simplify the problem by assuming that photoionization depends only on the metagalactic background light from \citet{haardt_radiative_2012}, and not from other location-dependent sources such as the central galaxy. This assumption is motivated by the evidence that local sources have a major effect mostly on the ionization state of the gas in the inner CGM $0.1-0.3R_{\rm vir}$  while gas outside this region receives a negligible fraction of the ionizing radiation from the galaxy \citep{sternberg_atomic_2002,sanderbeck_sources_2018}.

In the VELA simulations photoionization or collisional ionization is not directly simulated. Two kinds of metallicity are explicitly recorded: metallicity from SNIa (iron peak elements) and SNII (alpha elements). In order to analyse the ionization fraction of different ions we will follow a similar approach as the one in \citetalias{roca-fabrega_cgm_2019}. First we will get the total mass and density of the different species (e.g. $n_O$, $n_C$) by multiplying total SNIa or SNII metal mass by their respective abundances. It is important to mention that although in \citetalias{roca-fabrega_cgm_2019} we made the assumption that the Type II metals are entirely oxygen, and that Type Ia metals had no oxygen component, here we have relaxed this assumption by using a distribution of metals according to \citet{iwamoto_nucleosynthesis_1999}. However, as nearly 90 percent of all Type II supernovae ejecta is oxygen by mass, the effect of this change was minimal. The second step was to use the \textsc{cloudy} software \citep{ferland_cloudy_1998,ferland_2013_2013} to assign the corresponding ionization fraction to each ion species, based on the gas temperature, density, and on the redshift. Finally, to access the total population of any ion species, we need to multiply this fraction (e.g. $f_{\rm OVI}$) by the total amount of the individual nuclei of that species (e.g. $n_{\rm O}$), that is $n_{\rm OVI} = f_{\rm OVI}\cdot n_{\rm O}$. This procedure was implemented in the simulation analysis package \textsc{trident} \citep{hummels_trident_2016}, which is itself based in the more general \textsc{yt} \citep{turk_multi-code_2011} simulation analysis suite. To add these ion number densities in post-processing requires an assumption of local ionization equilibrium within each cell at each timestep. Note that this does not imply that we assume the gas to be in thermal or dynamical equilibrium. The gas can still be experiencing net cooling, or net heating due to feedback processes from the central galaxy. 

\begin{figure}
\includegraphics[trim={0.65cm 0.8cm 0.7cm 0.05cm}, clip, width=0.99\linewidth]{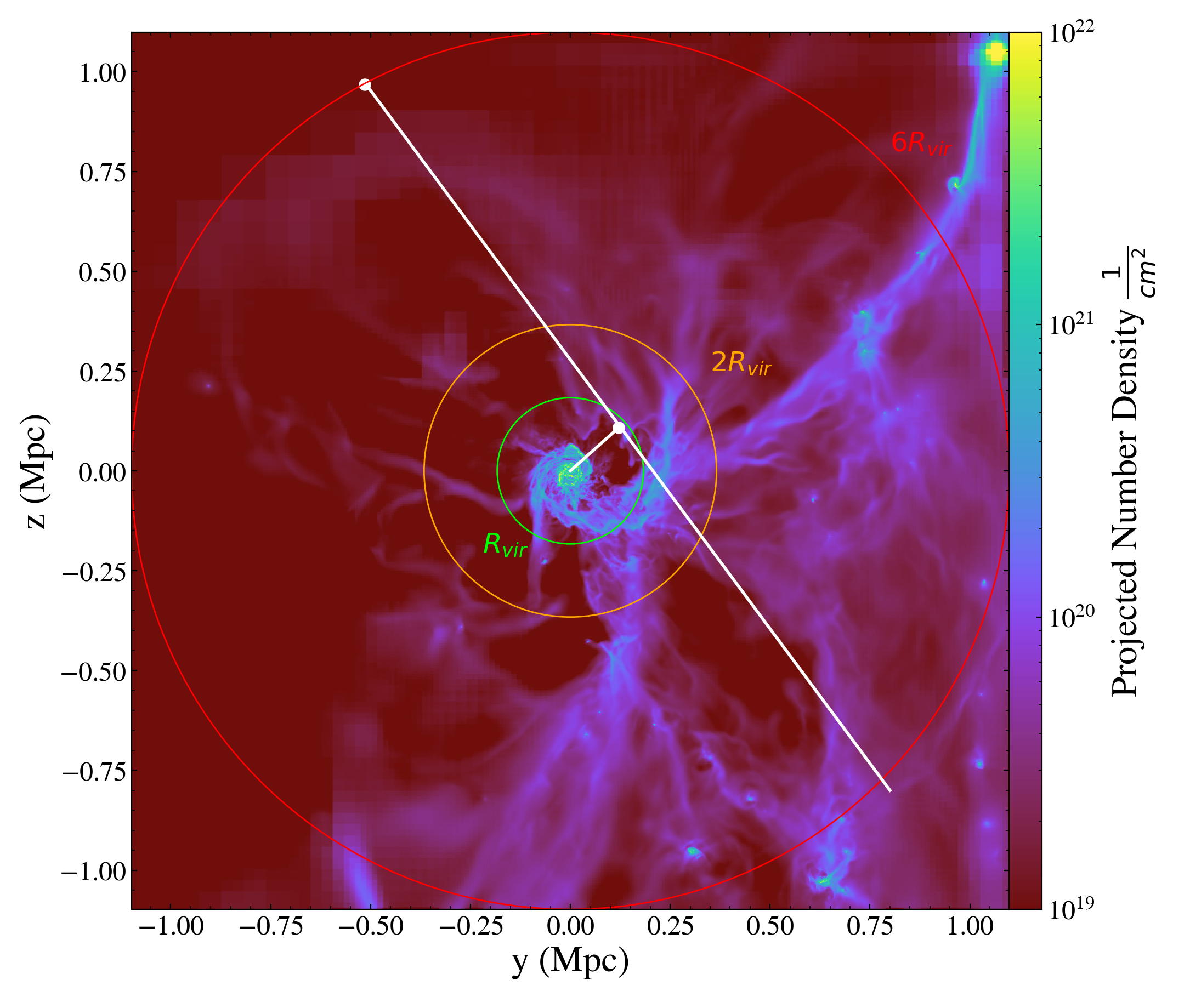}\par 
\caption{A sample sightline is generated from a random point on the outer sphere and directed to a midpoint at a specified impact parameter, denoted as white dots, and projected to twice that length. This is plotted over a projection of total gas density within the simulation, on the same snapshots and at the same angle as Figure \ref{fig:ci-frac}.}
\label{fig:sample}
\end{figure}

In order to emulate the absorption-line studies for direct comparison to observations, we create a large number of sightlines ($\sim400$) through each CGM. This procedure is similar to that of \citet{li_probing_2020}. The sightlines are defined via a startpoint and a midpoint. To choose the startpoint, we define a sphere at some maximum radius, outside of the simulation's `zoom-in region' (extending to $\sim 2 R_{\rm vir}$ from the center). This is for geometrical effect, and to make sure that no significant difference in path length appears between sightlines. It was confirmed by comparing results with all low-resolution (>15 kpc) cells removed and with them included that in no simulation did the gas outside the fiducial region have a significant impact on any results. We define this maximum radius $R$ to be at $6R_{\rm vir}$. We randomly choose one of a finite set of polar angles $\theta$ and a finite set of azimuthal angles $\phi$ according to a probability distribution scheme which distributes the startpoint uniformly across the surface of this sphere. The vector from the galaxy center to the startpoint is defined as normal to an `impact parameter' plane.
A midpoint is then selected from the plane at one of a discrete number of impact parameters, according to a probability distribution which gives a uniform point in the circle $r_\perp \leq 2R_{vir}$. A slight bias is introduced to give $r_\perp=0$ a non-zero chance of selection. However, this has a negligible effect on any results, and only affects our column densities for the few lines that go directly through the galaxy. This line is extended past the midpoint by a factor of 2. A visualization of a sightline generated by this algorithm is shown in Figure \ref{fig:sample}. 

This strategy is useful for several purposes. First, by choosing a finite set of sightlines, we can use the same statistical analysis methodology as used in observational studies. In particular, in Section \ref{sec:observations}, we can emulate the inverse-Abel transformation used in \citetalias{stern_does_2018}. Second, we save a significant amount of information within each sightline, allowing us to track correlations between ions within sightlines, and the state of gas within the sightlines (see Section \ref{sec:definitions}), instead of losing it in continual averaging. 

\section{Collisional and Photo Ionization}\label{sec:definitions}
In this section we present a physically motivated definition of `Collisionally Ionized' and `Photoionized' gas as distinct states which coexist throughout the CGM. We do this both specifically for OVI, as well as for all other ion species. We will refer to these states hereafter as CI and PI, respectively. We will also refer often to the following temperature states, in accordance with \citetalias{roca-fabrega_cgm_2019}, \citetalias{stern_universal_2016}, and \citet{faerman_massive_2017}.
\begin{itemize}
\item Cold gas: T $<$ 10$^{3.8}$K
\item Cool gas: 10$^{3.8}$K $<$ T $<$ 10$^{4.5}$K
\item Warm-hot gas: 10$^{4.5}$K $<$ T $<$ 10$^{6.5}$K
\item Hot gas: T $>$ 10$^{6.5}$K
\end{itemize}

In \citetalias{stern_does_2018}, two scenarios are outlined for CGM gas in phase space, which generated OVI either the higher-density, hotter peak (CI) or the lower-density, cooler peak (PI) However it was not clear how to classify gas either far from these peaks or near both, where the OVI fraction is nonnegligible but clearly depends sensitively on both temperature and density. We will present a different definition here, which agrees qualitatively with that definition and the procedure in \citetalias{roca-fabrega_cgm_2019}, but has some differences. It also bears some resemblance to the definition in \citet{faerman_massive_2020}, where the oxygen ion distribution found under a pure CIE definition was compared to the distribution from the distribution including PI, and then identified critical densities below which photoionization becomes important.

We will define these two states (CI and PI) graphically, using the data from \textsc{cloudy} at redshift 1, assuming a uniform \citet{haardt_radiative_2012} ionizing background. At a given temperature, the distribution of ions for a single atomic species is a function of density. At sufficiently high density, for each ion the ionization fraction either decreases to 0, converges to a stable nonzero fraction, or (for the neutral atom at low temperatures) increases to 1. At sufficiently low density, ionization fractions drop to 0 for all ions except for `fully ionized' states. See Figure \ref{fig:ionfractions} for examples of both behaviors. There are three `characteristic' shapes to these graphs, and so at each temperature T, each ion can be categorized as one of the following: 

\begin{figure}
   \centering
      \includegraphics[trim={1.2cm 0.5cm 1.9cm 2.0cm}, clip, width=0.99\linewidth]{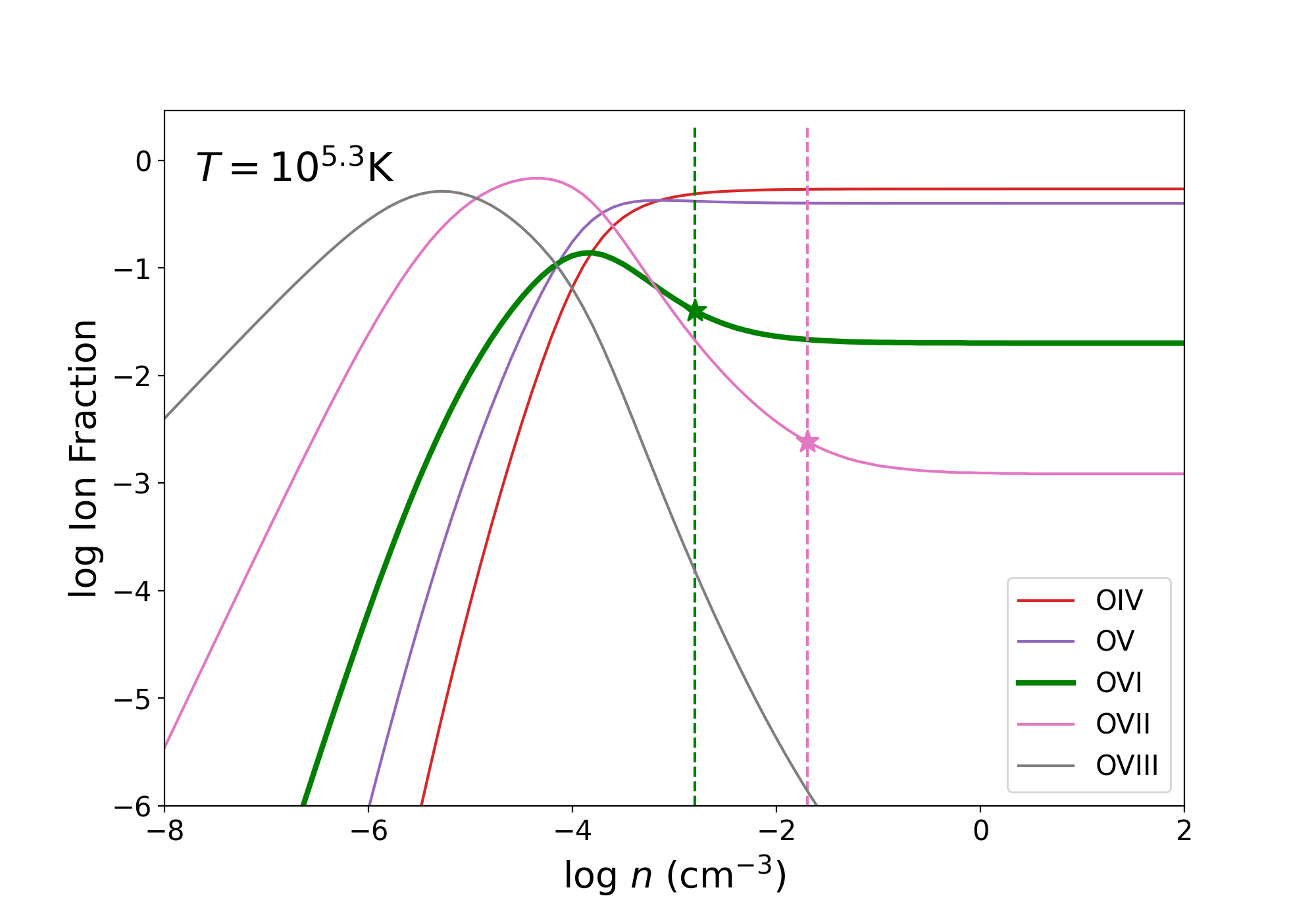}
   \caption{Examples of the three characteristic shapes of ion fraction distributions, at a particular fixed temperature ($T=10^{5.3} {\rm K}$). Here, O IV and O V are fully CI, OVI and OVII have transition points at $\log n = -2.5$ and $-1.8$, respectively, and OVIII is fully PI. This can be seen by the end-behavior at high density. The stars indicate the same points as in Figure \ref{fig:cipidef}.}
   \label{fig:ionfractions}
\end{figure}
\begin{itemize}
    \item `fully CI': flat after some high density, falls directly to 0 without any significant increase at low density (see OIV, red line in Figure \ref{fig:ionfractions}). Clearly photoionization affects this state at low density, but the critical element is that photoionization only \textit{destroys} this state, and does not create it. So we will claim this ion's creation is density-independent, and therefore only depends on temperature, or in other words, is CI. 
    \item `fully PI': Does not stabilize at high density, but rather decays to 0 after reaching a maximum at some intermediate density (see OVIII, grey line in Figure \ref{fig:ionfractions}). Since the ionization fraction is always a strong function of density, this gas is PI. 
    \item `transitionary': Stabilizes at high density, but also contains a maximum which is higher than that stable fraction (see OVI and OVII, green and pink lines in Figure \ref{fig:ionfractions}). We will define a `transition density' to be the density at which the ion fraction is exactly twice the stable CI fraction, and if the maximum is not this high, the ion is considered `fully CI' because while there is a non negligible PI fraction, it is never dominant (see OV, purple line in Figure \ref{fig:ionfractions}).
\end{itemize}
So, for each temperature, each ion can be characterized as PI, CI, or it may be at a transitionary temperature, and then it will be PI on the left and CI on the right of a transition density. The pair of a transitionary temperature and the associated transition density at that temperature will be called a \textit{transition point}. Iterating over all temperatures from T=$10^{2.5}$ K to T=$10^{7.5}$ K in steps of 0.1 dex, we found that each species starts out as PI at low temperatures, has transitionary temperatures for several consecutive values of $T$, each time decreasing the transition density $n$, and then becomes fully CI at sufficiently high temperatures. In Figure \ref{fig:cipidef}, we plot the lines of transition from PI to CI in $T-n$ space. Considering the above discussion, we extend the lines on the right to $\log n=+\infty$ from the lowest-temperature transition point, and on the left to $\log n=-\infty$ from the highest-temperature transition point. The two transition points from Figure \ref{fig:ionfractions} are marked again with stars in Figure \ref{fig:cipidef}. Note that in general ion species do not have the same number of transitionary temperatures, and in fact OII (orange line in Figure \ref{fig:cipidef}) has none. This means it is `fully PI' below T=10$^{4.2}$ K and `fully CI' above. We speculate that with a higher resolution in T space, a narrow temperature range around 10$^{4.2}$ K would be found where transitionary temperatures exist for OII. It is also true that there are regions of $n-T$ space in which an ion will be classified as PI or CI by this definition even though that ion will have a fraction of approximately zero there. This naturally will have no effect on the distribution of the ion into the two states, as insignificant regions of the graph will have negligible contributions. 
We see that in this graph, low ions have their transition from CI to PI in or near cool gas, mid-range ions have transitions in the middle range of warm-hot gas, and high ions are CI only in hot gas and PI in warm-hot gas\footnote{A full table for all ions mentioned in this paper, at redshifts from 1 to 4, is available online.}. We note that this implies that whether an ion should be considered `generally' CI or PI throughout the CGM is a more complicated story than usually assumed in the literature. For example, it is accurate to say that low ions (MgII, SiII, OII etc) only exist in cool or cold gas, but for them the CI-PI cutoff is also located within cool gas, so it would be inaccurate to say that those low-ion states are necessarily photoionized (see OII, orange, in Figure \ref{fig:cipidef}). A similar statement would be true about OVII and OVIII: they certainly can only exist within warm-hot/hot gas, but this does not imply they are entirely collisionally ionized. 
\begin{figure}
   \centering
      \includegraphics[trim={1.2cm 0.5cm 1.9cm 2.0cm}, clip, width=0.99\linewidth]{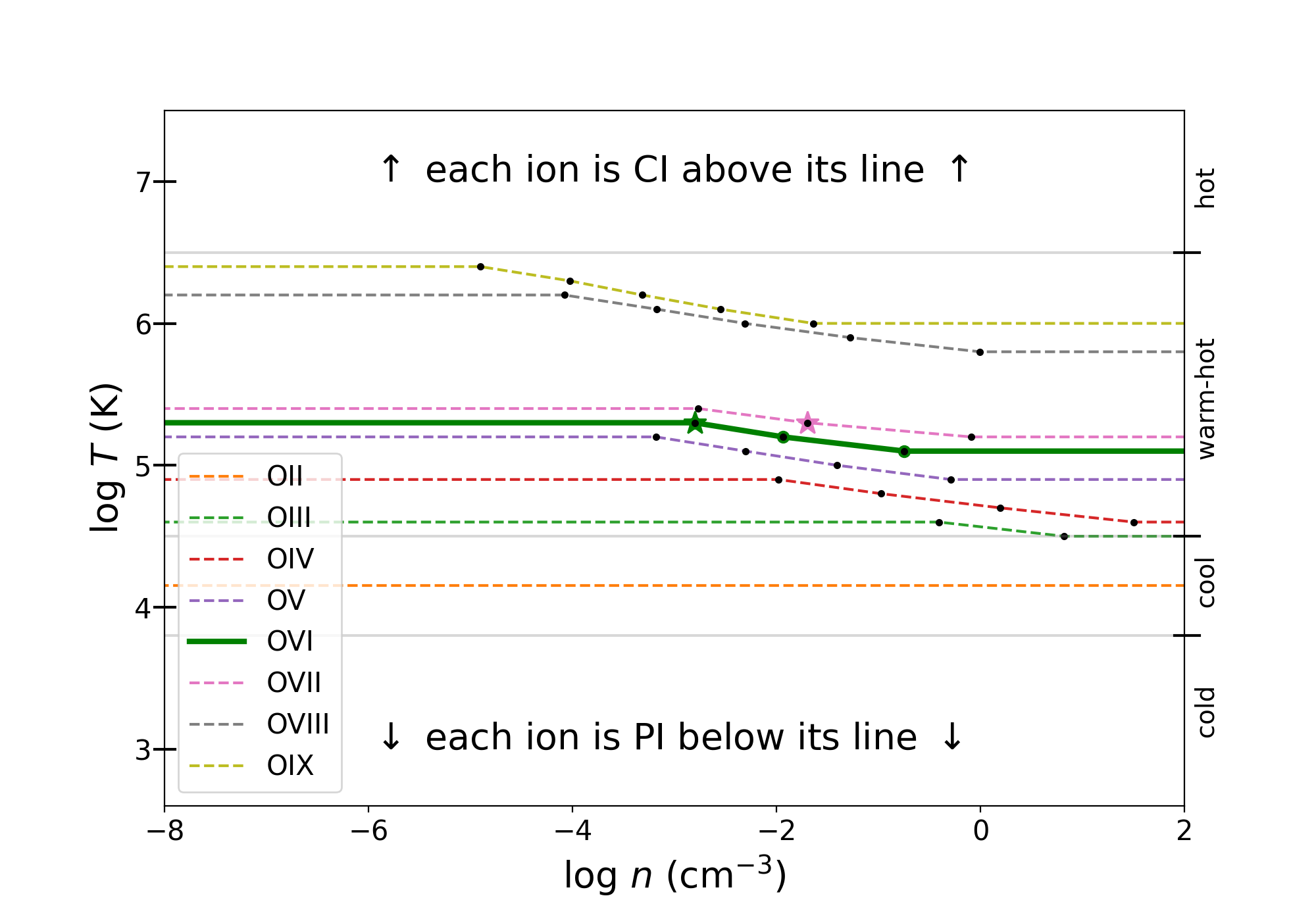}
   \caption{Using the density-dependent ion fraction curves (Figure \ref{fig:ionfractions}) and plotting transition points (dots), we can describe all ions as PI or CI according to their position in $T-\rho$ phase space. This definition is independent of the ion fraction at that position, which might be negligible. Each dot is a CI-PI cutoff at a particular temperature and density, with the cutoffs from Figure \ref{fig:ionfractions} marked with green and pink stars.} 
   \label{fig:cipidef}
\end{figure}
We can visualize the results of these definitions using a binary field in \textsc{yt}. We define a field CI\_OVI to be 1 if OVI is CI-dominated in that specific cell, 0 otherwise, and PI\_OVI to be the opposite. Multiplying this field by the actual OVI density allows us to differentiate the two populations of OVI, and similar methods can be applied to other ions.

\section{Distributions of PI and CI OVI Gas}\label{sec:sec4overall}
We now analyse the actual spatial distribution of OVI within the CGM of the VELA simulations. Unless otherwise noted, we will refer to the gas outside 0.1 $R_{\rm vir}$ and within $R_{\rm vir}$ as the CGM, though in fact, recent studies \citep{wilde_cgm2_2020} have shown that $\Rv$ is not really a `physical' boundary to the CGM and probably underestimates the dynamical conditions of the CGM. However considering our decreased resolution outside the virial radius (Figure \ref{fig:resolution}) and the fact many analytic models use $\Rv$ as a starting point (see below, Section \ref{sec:streams_theory}), we will continue to use this definition. We focus here on VELA07 at redshift 1, but other VELA galaxies are similar at this redshift. This galaxy is plotted in a plane which is approximately face-on, with the $x$ axis being at a 25 degree angle from the galaxy angular momentum, which was calculated in \citet{mandelker_giant_2017}, and is a large spiral galaxy. Other views of the same galaxy, including the overall distribution of gas and stars, can be seen in Figures 1, 4, 19, and D2 of \citet{Dekel20b}\footnote{Images of all of the VELA galaxies at a variety of angles, as they would appear using HST, JWST, or other instruments, are available at \url{https://archive.stsci.edu/prepds/vela/}}. We will start with the distribution in 3D space, and then look within projections at the fractions of CI and PI gas. We will continue to use the terminology from \citetalias{stern_does_2018} and call the radius of the median OVI ion $R_{\rm OVI}$, or the `half OVI radius'. We will show that within the simulation, $R_{\rm OVI}$ is indeed outside half of the virial radius, as suggested by COS-Halos \citepalias{stern_does_2018}. Furthermore, because OVI extends past the virial radius and because of the concavity of the deprojected OVI profile (see Section \ref{sec:observations}), we find that $R_{\rm OVI}$ is likely even larger than suggested in \citetalias{stern_does_2018}. 

\subsection{OVI Distribution in 3D space}\label{sec:3Ddistribution}

\begin{figure*}
\begin{multicols}{2}
    \includegraphics[trim={0.65cm 0.8cm 0.7cm 0.05cm}, clip, width=0.99\linewidth]{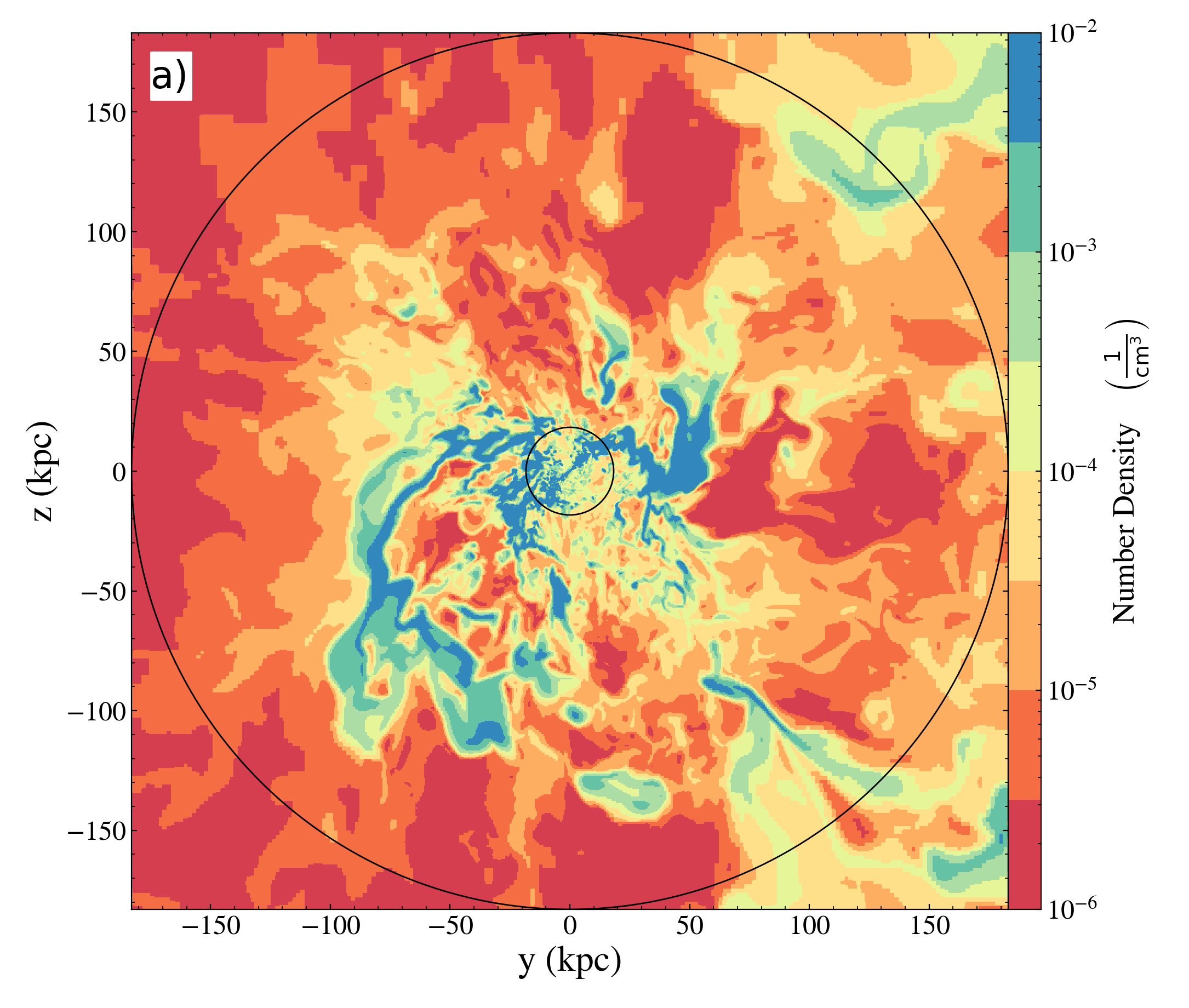}\par
    \includegraphics[trim={0.7cm 0.8cm 0.7cm 0.05cm}, clip, width=0.99\linewidth]{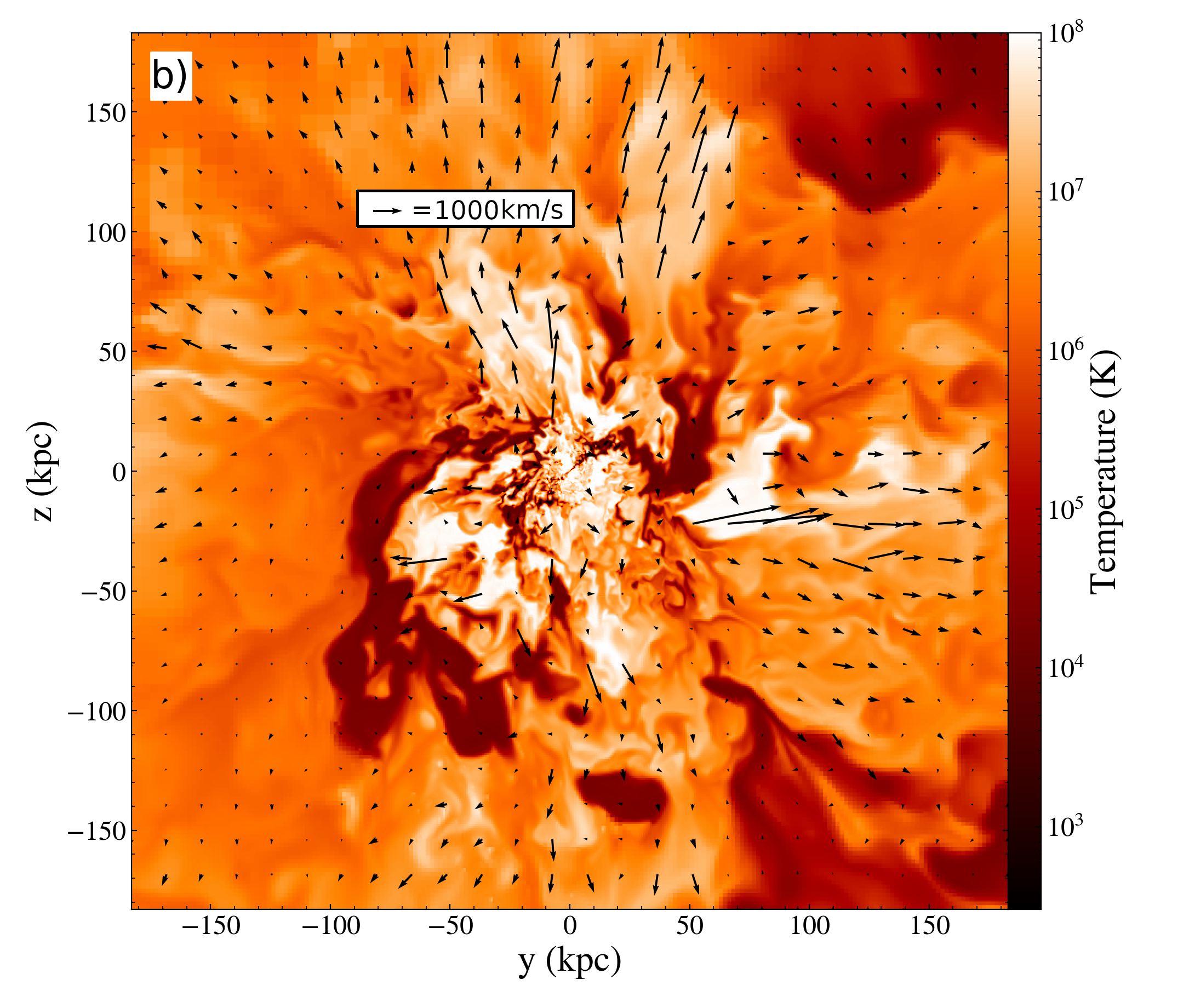}\par
\end{multicols}
\begin{multicols}{2}
    \includegraphics[trim={0.65cm 0.8cm 0.7cm 0.05cm}, clip, width=0.99\linewidth]{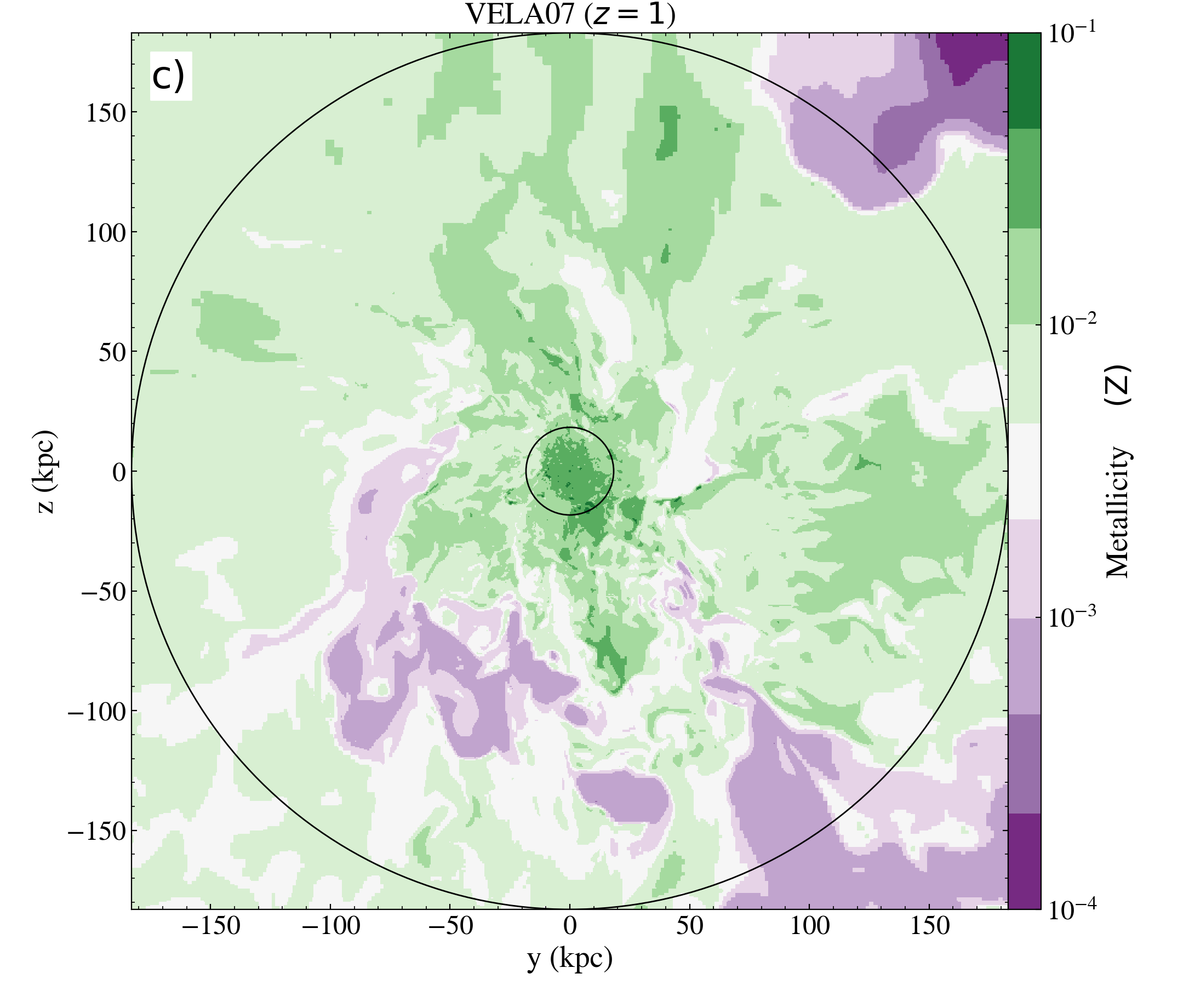}\par
    \includegraphics[trim={0.65cm 0.8cm 0.7cm 0.05cm}, clip, width=0.99\linewidth]{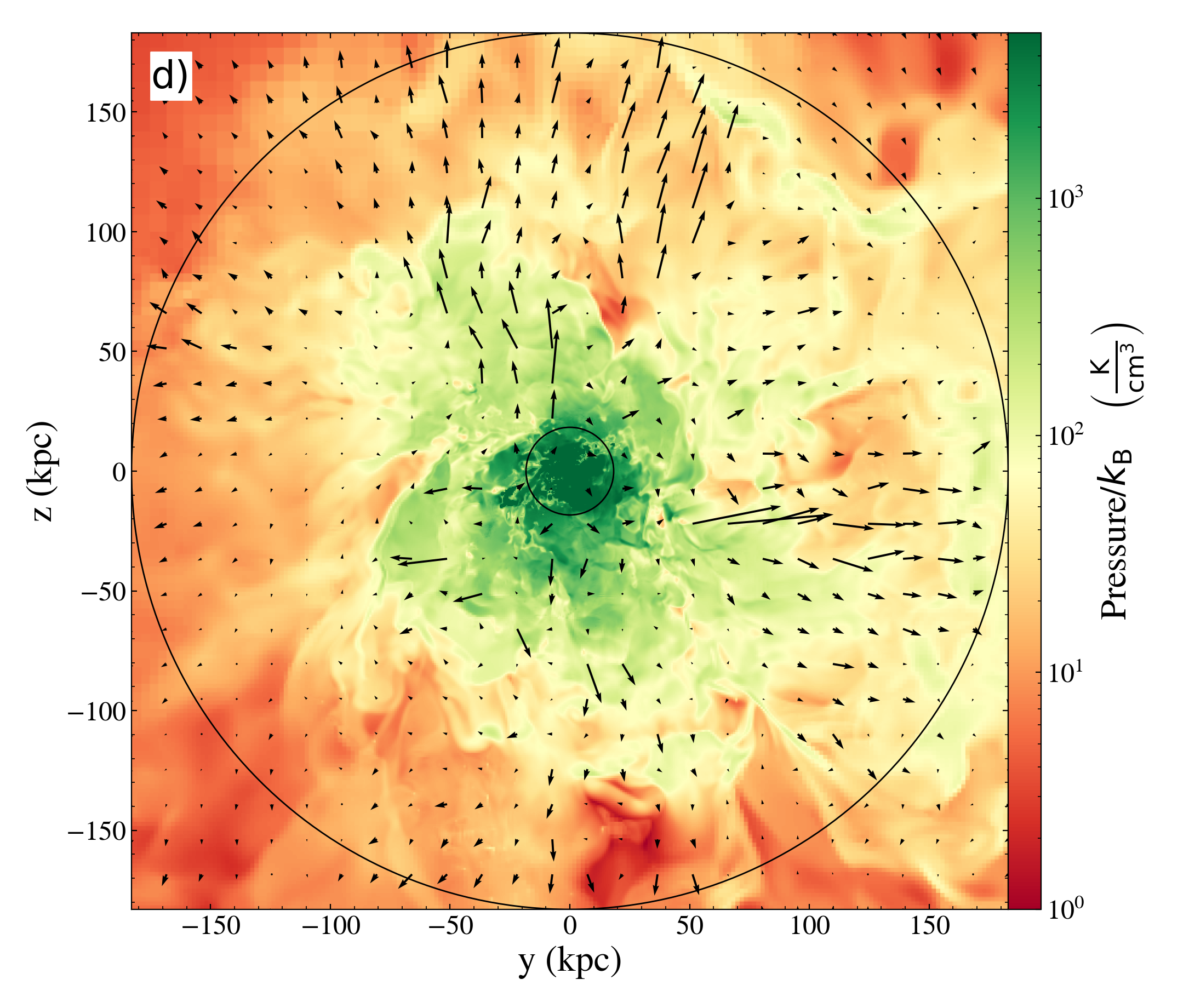}\par
\end{multicols}
\caption{2D slice of VELA07 at redshift 1, approximately face-on. The smaller and larger black circles in each image represent $0.1R_{\rm vir}$ and $R_{\rm vir}$. \textbf{(a):} Number density of all gas within the simulation (i.e. hydrogen density). The clouds shown in this slice form two continuous streams, at the bottom and top right, but since the image is only a thin slice, it appears patchy as it moves slightly in and out of the plane \textbf{(b):}  Temperature of gas within the slice. Overplotted is gas velocity within the cell, with respect to the galaxy center. \textbf{(c):} Gas metallicity of the simulation. The same streams are highlighted in each image (high density, low temperature, low metallicity). \textbf{(d):} Pressure of gas within the simulation, defined in units of number density times temperature. Arrows are the same as in b. The streams are seen here to be in approximate pressure equilibrium with their surroundings.}
\label{fig:GASPROPSslice}
\end{figure*}

\begin{figure*}
\begin{multicols}{2}
    \includegraphics[trim={0.65cm 0.8cm 0.7cm 0.05cm}, clip, width=0.99\linewidth]{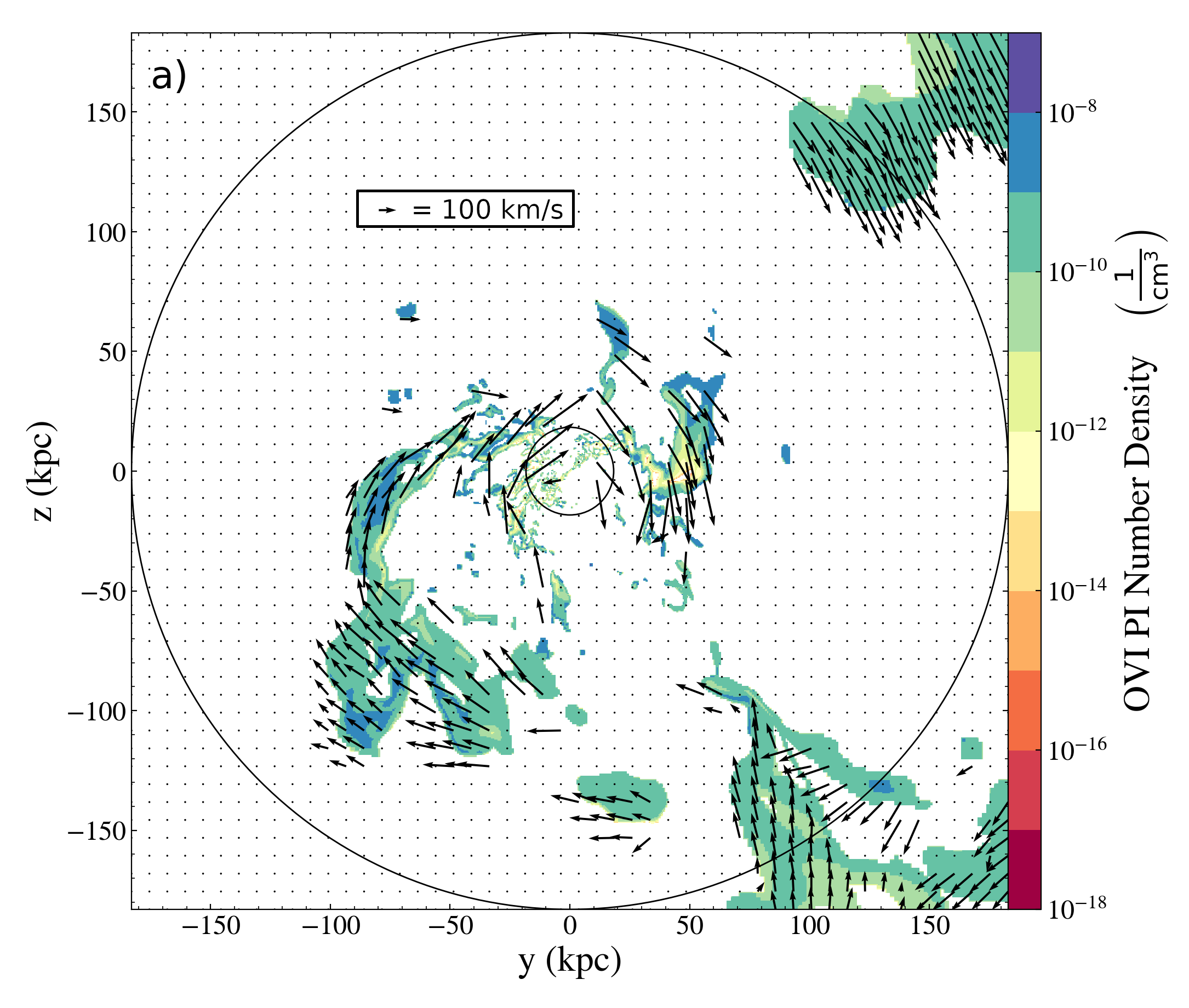}\par 
    \includegraphics[trim={0.65cm 0.8cm 0.7cm 0.05cm}, clip, width=0.99\linewidth]{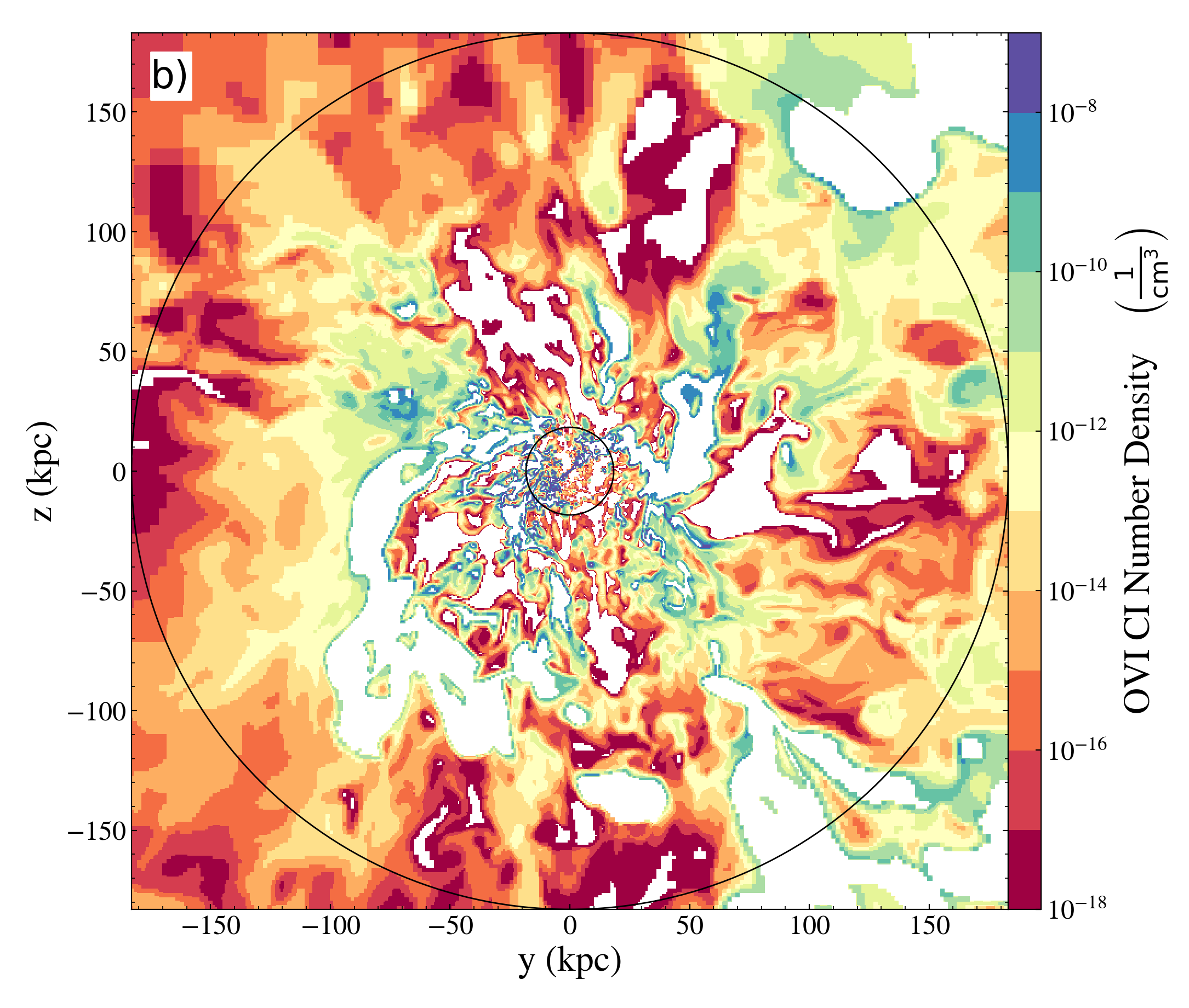}\par
    \end{multicols}
\begin{multicols}{2}
    \includegraphics[trim={0.65cm 0.8cm 0.7cm 0.05cm}, clip, width=0.99\linewidth]{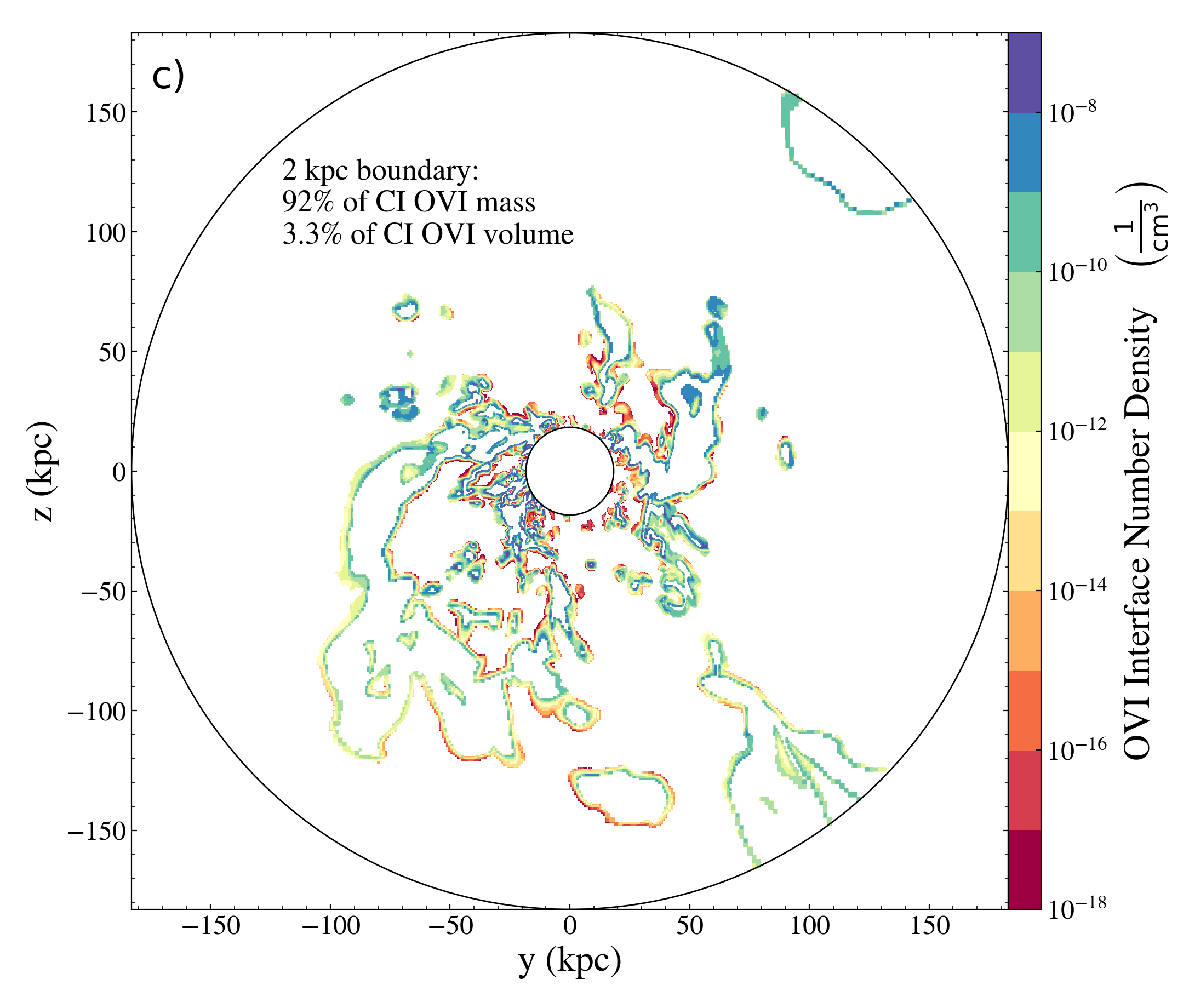}\par
    \includegraphics[trim={0.65cm 0.8cm 0.7cm 0.05cm}, clip, width=0.99\linewidth]{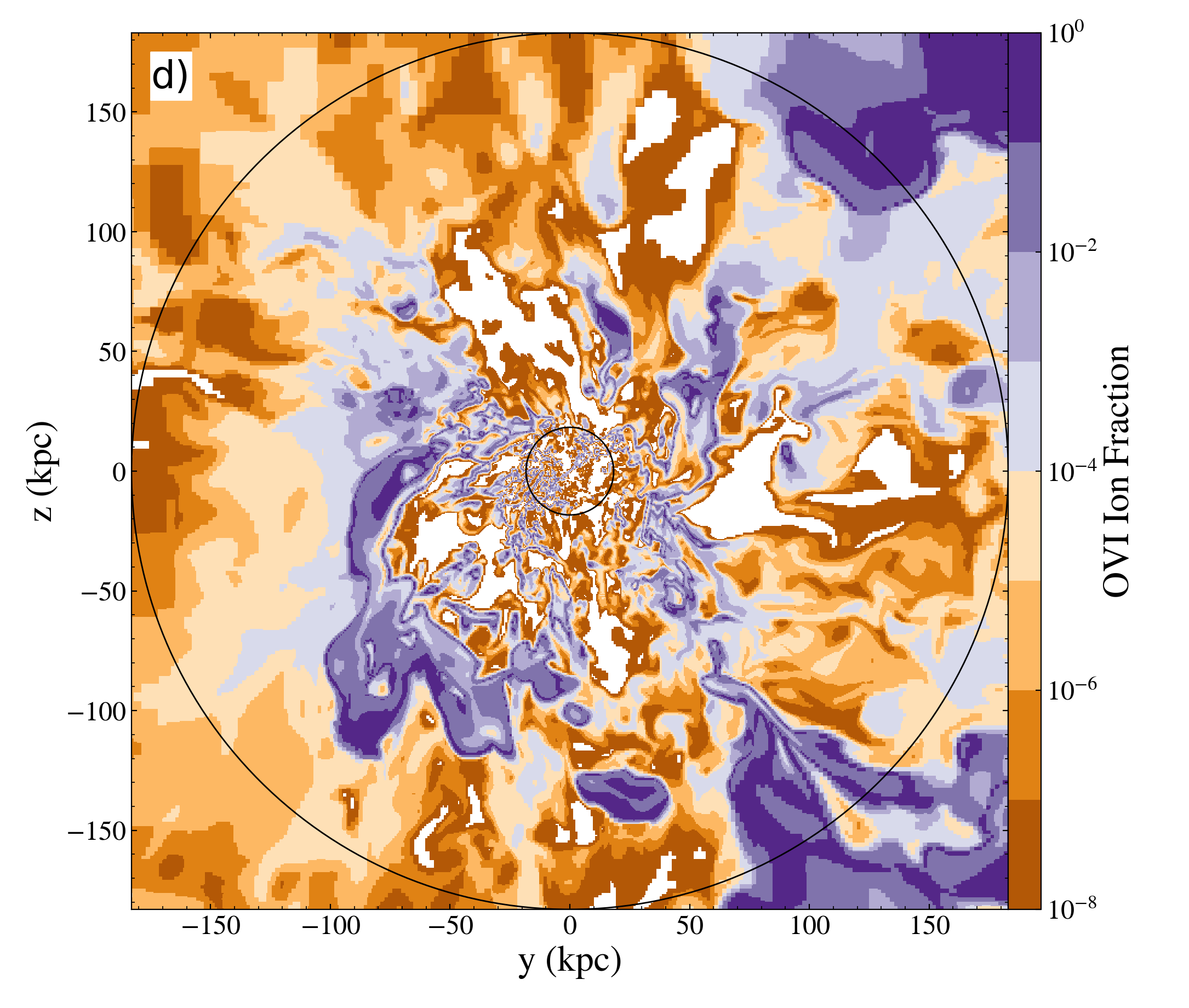}\par
\end{multicols}
\caption{OVI properties within the same 2D slice of VELA07 shown in Figure \ref{fig:GASPROPSslice}. \textbf{(a):} OVI number density for only gas defined as PI. Overplotted is gas velocity within the cell, with respect to the galaxy center. Only cells which are classified as PI have their velocities shown here. The scale of the arrows in this graph is smaller than in Figure \ref{fig:GASPROPSslice}b. \textbf{(b):} OVI number density, for gas defined as CI. The two populations are defined to be mutually exclusive. \textbf{(c):} OVI number density restricted only to the interface and to within 0.1 - 1.0 $R_{\rm vir}$, where the interface is defined as CI cells within 2 kpc of a PI cloud. \textbf{(d):} ion fraction (fraction of oxygen nuclei ionized to OVI).}
\label{fig:OVIslice}
\end{figure*}

In Figures \ref{fig:GASPROPSslice} and \ref{fig:OVIslice} we analyse a 2D slice through a representative simulation (VELA07 at $z=1$). As a 2D slice it has zero thickness, however since the simulation has finite resolution the effective thickness is the resolution of the simulation (see Figure \ref{fig:resolution}).
Figure \ref{fig:GASPROPSslice} shows several macroscopic properties of gas within the simulation. Here the features visible in this plane are the inflowing streams of cool, high-density gas (see Figure \ref{fig:OVIslice} for evidence of inflows) and the hot medium surrounding them. We will call these structures the `streams' and `bulk', respectively. These are the streams of baryonic matter necessary to feed star formation, and have been studied before in VELA (\citealp{zolotov_compaction_2015}; \citetalias{roca-fabrega_cgm_2019}), and in similar simulations \citep{ceverino_high-redshift_2010,danovich_four_2015}. While these streams look discontinuous, they only appear so due to minor fluctuations moving them outside the plane of the slice. They are part of the same counterclockwise-spiraling dense streams visible in the projection of Figure \ref{fig:sample}. Overplotting the in-plane velocity, we see that within the hot gas are fast outflows with velocities $\sim$1000 km s$^{-1}$. On this scale, the inflowing speed of the cool gas is not visible. As noted in \citetalias{roca-fabrega_cgm_2019}, the metallicity of these streams is substantially lower than the surrounding hot gas, which indicates that they are infalling from the metal-poor IGM and not primarily cooling out of the halo gas. However, the increased density in these regions more than makes up for their lower metallicity, so we expect them to be detectable in metals. 
In fact they are essentially traced out by OVI (see below). Finally, we see in the bottom right panel (Figure \ref{fig:GASPROPSslice}d) that none of these structures are reflected in the pressure diagram, and in fact pressure is almost spherically symmetric, with a maximum in the central galaxy. So, `overall' the cool inflows are in approximate pressure equilibrium with the rest of the galaxy. However, on closer inspection we find that on average throughout the inflowing streams, pressure is at least 50 percent lower than the bulk, and on the inner halo a factor of 2-3 lower, so while it is true that pressure differentials are not the primary feature of these cool inflowing streams, their lower pressure does allow them to fall towards the centre.

Focusing on OVI within the same slice, Figure \ref{fig:OVIslice}a and \ref{fig:OVIslice}b show the separation of the slice into PI cells (left) and CI cells (right) according to the definition in Section \ref{sec:definitions} (see green line in Figure \ref{fig:cipidef}). Since they are defined to be mutually exclusive, the filled cells in the left panel appear as white space in the right panel, and vice versa. We see from this that CI gas fills the majority of the volume of this slice (more quantitative results, which do not depend on the specific snapshot and slice orientation, can be found in Table \ref{tab:2}) and PI gas is found only inside the cool inflows. This follows from the temperature distribution since, as expected from Section \ref{sec:definitions} the PI-CI cutoff is nearly equivalent to a temperature cutoff.  Also marked in the left plot is the velocity of the PI cells within the slice. Since the OVI ions are added to the simulation in post-processing, the velocity is the overall gas velocity in that region, not the separate velocity of only OVI. This shows that PI OVI clouds, and therefore also the streams shown in Figure \ref{fig:GASPROPSslice}, are generally inflowing and rotating, with a characteristic velocity of $\sim$ 100 km s$^{-1}$, significantly slower than the outflows. It is contained within filaments which become smaller in cross-section as they spiral towards the central galaxy. 

Next we analyse the distribution of OVI by mass (instead of volume as in the previous paragraph) between the two states. It is clear from the top panels of Figure \ref{fig:OVIslice}(a and b) that the OVI number density is higher in the PI clouds than in the CI bulk, but since the clouds fill only a small fraction of the CGM, it is not \textit{a priori} clear which phase would dominate in either sightlines or in the CGM overall. In \citetalias{roca-fabrega_cgm_2019} it was found that this depended strongly on redshift and galaxy mass. All galaxies at high redshift were CI-dominated and become more PI-heavy with decreasing reshift, eventually diverging by mass, with larger galaxies approaching CI-domination again at low redshift, and smaller galaxies remaining PI-dominated to redshift 0. We find here (Table \ref{tab:2}, column 3) that the PI gas contains about 2/3 of the OVI mass, showing OVI is primarily found in the CGM in cooler, lower-metallicity gas, but that a nonnegligible fraction remains CI.

Before we can address which phase will dominate within sightlines, there is one other important feature of Figure \ref{fig:OVIslice}. While the number density of the CI bulk is significantly lower than the PI clouds, within the CI slice (Figure \ref{fig:OVIslice}b) there are small regions of high number density. These are present only along the edges of the PI clouds themselves. To get a more quantitative understanding of this `interface layer', we used a KDTree algorithm to query each CI gas cell within 0.1-1.0 $R_{\rm vir}$ as to its nearest PI neighbor. We define the interface (for now, see Section \ref{sec:model_2} for more details) as CI cells which had any PI neighbor within 2 kpc. In the bottom left panel, (Figure \ref{fig:OVIslice}c) we show only the interface cells as defined above. 
These cells consist of $\sim 92$ percent of the CI OVI mass within the virial radius, while occupying only $\sim 3.3$ percent of the CI volume. Therefore, in addition to dividing the CGM into CI and PI components, we believe it is useful to further divide the CI gas into two phases: interface and bulk. An interface layer like the one shown in Figure \ref{fig:OVIslice} is often found surrounding cold dense gas flowing through a hot and diffuse medium \citep{gronke_growth_2018,gronke_how_2020,ji_simulations_2019,li_fate_2019,mandelker_instability_2020,fielding_multiphase_2020}. In Section \ref{sec:model_2} we will present a model for the physical origin and properties of the interface layers found in our simulations. 

The fraction of gas in each phase, calculated using both a 2 kpc and a 1 kpc boundary, is shown in Table \ref{tab:2}. A complicating factor, which is outside the scope of this paper to address, is that within these boundary layers, gas is unlikely to be in ionization equilibrium \citep{begelman_turbulent_1990,slavin_turbulent_1993,kwak_numerical_2010,kwak_simulations_2011,oppenheimer_bimodality_2016} and so it is possible that the mass distribution of CI gas in the two phases will differ significantly from that presented here. In particular, it was found in \citet{ji_simulations_2019} that nonequilibrium ionization can increase the column densities of OVI by a factor of $\sim (2-3)$ within turbulent interface layers. 

In this simulation, outflowing warm gas is generally too hot to have a significant OVI contribution, making the bulk of the volume CI but negligible in OVI outside the inner halo ($\sim 0.3 R_{\rm vir}$). We are not claiming that the total gas density in these warm-hot outflows is extremely low, but only their OVI number density. This could be due to a low value of any of the contributing factors of ion fraction, density, or metallicity. We find in these simulations that it is primarily the ion fraction which causes this bulk volume to be negligible in OVI, compared to 
both the interface and the cool streams (see \se{other_ions}). Outflows will never have low metallicity, as the outflows are driven by supernova winds \citep[see previous ART simulations, e.g.][]{ceverino_role_2009}. Additionally, as seen in Figure \ref{fig:GASPROPSslice}, the high density and low metallicity inside the streams mostly cancel one another. However, both of those effects are much smaller than the the ion fraction dependence (Figure \ref{fig:OVIslice}d). 

\begin{table*}
\begin{centering}
\begin{tabular} {|l|c|c| c| c| c| c| c| c| c|} 
 \hline
VELA & edge  & OVI PI & OVI  CI& OVI CI  & OVI PI  & OVI  CI & OVI CI & CI edge \\
 ($z=1$)& width & mass \%  & edge mass \% & bulk mass \%  & vol \%  & edge vol \% & bulk vol \%& mass \%  \\
 \hline
 {\bf V07} & {\bf 1 kpc} & {\bf 63} & {\bf 30} & {\bf 7}  & {\bf 5} & {\bf 1} & {\bf 94}& {\bf 81} \\
 V08 & 1 kpc & 48 & 31 & 20  & 9 & 2 & 89& 61 \\
 V10 & 1 kpc & 68 & 21 & 11  & 18 & 2 & 80& 66 \\
 V21 & 1 kpc & 40 & 39 & 20  & 6 & 1 & 93& 66 \\
 V22 & 1 kpc & 80 & 16 & 4  & 4 & 1 & 95& 80 \\
 V29 & 1 kpc & 97 & 1 & 1  & 23 & 1 & 77& 50 \\
 \hline
 {\bf Stacked} & {\bf 1 kpc} & {\bf 62} & {\bf 26} & {\bf 12}  & {\bf 10} & {\bf 1} & {\bf 89}& {\bf 68} \\
 \hline
 {\bf V07} & {\bf 2 kpc} & {\bf 63} & {\bf 35} & {\bf 3}  & {\bf 5} & {\bf 3} & {\bf 91}& {\bf 92} \\
 V08 & 2 kpc & 48 & 43 & 8  & 9 & 7 & 84& 84 \\
 V10 & 2 kpc & 68 & 27 & 5  & 18 & 6 & 76& 84 \\
 V21 & 2 kpc & 40 & 51 & 9  & 6 & 4 & 90& 85 \\
 V22 & 2 kpc & 80 & 18 & 2  & 4 & 2 & 94& 90 \\
 V29 & 2 kpc & 97 & 2 & 1  & 23 & 4 & 77& 66 \\
 \hline
 {\bf Stacked} & {\bf 2 kpc} & {\bf 62} & {\bf 33} & {\bf 5}  & {\bf 10} & {\bf 4} & {\bf 86}& {\bf 87} \\
 \hline
 \end{tabular}
  \caption{Distribution of CGM gas (within $0.1-1.0 R_{\rm vir}$) within the selected VELA snapshots at redshift $z=1$. Columns 3, 4, and 5 are the mass distribution of OVI into PI gas, CI edge gas, and CI bulk gas (so they should sum to 100 percent, ignoring truncation errors). Columns 6, 7, and 8 are the volume distribution into the same categories. Column 9 is the percentage of CI gas by mass within the edge. The same analysis is repeated with an assumption of a 1 kpc edge and a 2 kpc edge.}
\label{tab:2}
 \end{centering}
\end{table*}

The distribution of OVI into the three categories within $0.1-1.0R_{\rm vir}$ for each VELA simulation, and the `stacked' results (the sum of the total values from each category) are shown in Table \ref{tab:2}. Here we see that in each simulation, photoionized OVI consists of $\sim (40-90)$ percent of all OVI by mass, with an average of approximately $62$ percent, and the CI gas is mostly concentrated within the interface. This parallels the findings of \citetalias{roca-fabrega_cgm_2019}, where it was found that cool gas dominates the CGM by mass, and warm-hot gas dominates the CGM by volume. Taking PI and CI OVI to be analogues of `cool' and `warm-hot' gas respectively, we see that OVI has a similar distribution. 
An interface of 1 kpc contains about two-thirds of CI gas, while a 2 kpc interface contains almost $90$ percent of CI gas. From a volume perspective, the CI bulk occupies the vast majority ($\sim 90$ percent) of the CGM, and this does not change appreciably when we consider a 2 kpc boundary layer instead of a 1 kpc boundary. There are effects which both underestimate and overestimate the amount of gas in these interfaces. In the outer parts of the CGM, the resolution is in fact worse than 1-2 kpc, and so even cells adjacent to PI clouds might not register as interface cells, underestimating that value. On the other hand, in the inner part of the CGM the resolution is much better and the 1-2 kpc cutoff might include some gas which is not dynamically `boundary layer gas' (see \se{model_2}). 

\subsection{OVI Within Sightlines}
\label{sec:sightlines}

While we see in Table \ref{tab:2} that by mass, the majority of all OVI within the CGM is PI, the projection of OVI through sightlines will distort the distribution, biasing the observed OVI gas towards the outer halo compared to the impact parameter. This is because the impact parameter of gas is the minimal distance of gas along the sightline, and all gas interacted with will be at that distance or farther. We saw in \citetalias{roca-fabrega_cgm_2019} that, regardless of galaxy mass and redshift, the outer halo was generally more PI than the inner halo, so we should expect the average sightline to be more PI than the gas distribution itself. However, the small volume filling factor could conceivably lead to a majority of sightlines not hitting any PI gas whatsoever. 

\begin{figure}
\includegraphics[trim={1.2cm 0.5cm 1.9cm 2.0cm}, clip, width=0.99\linewidth]{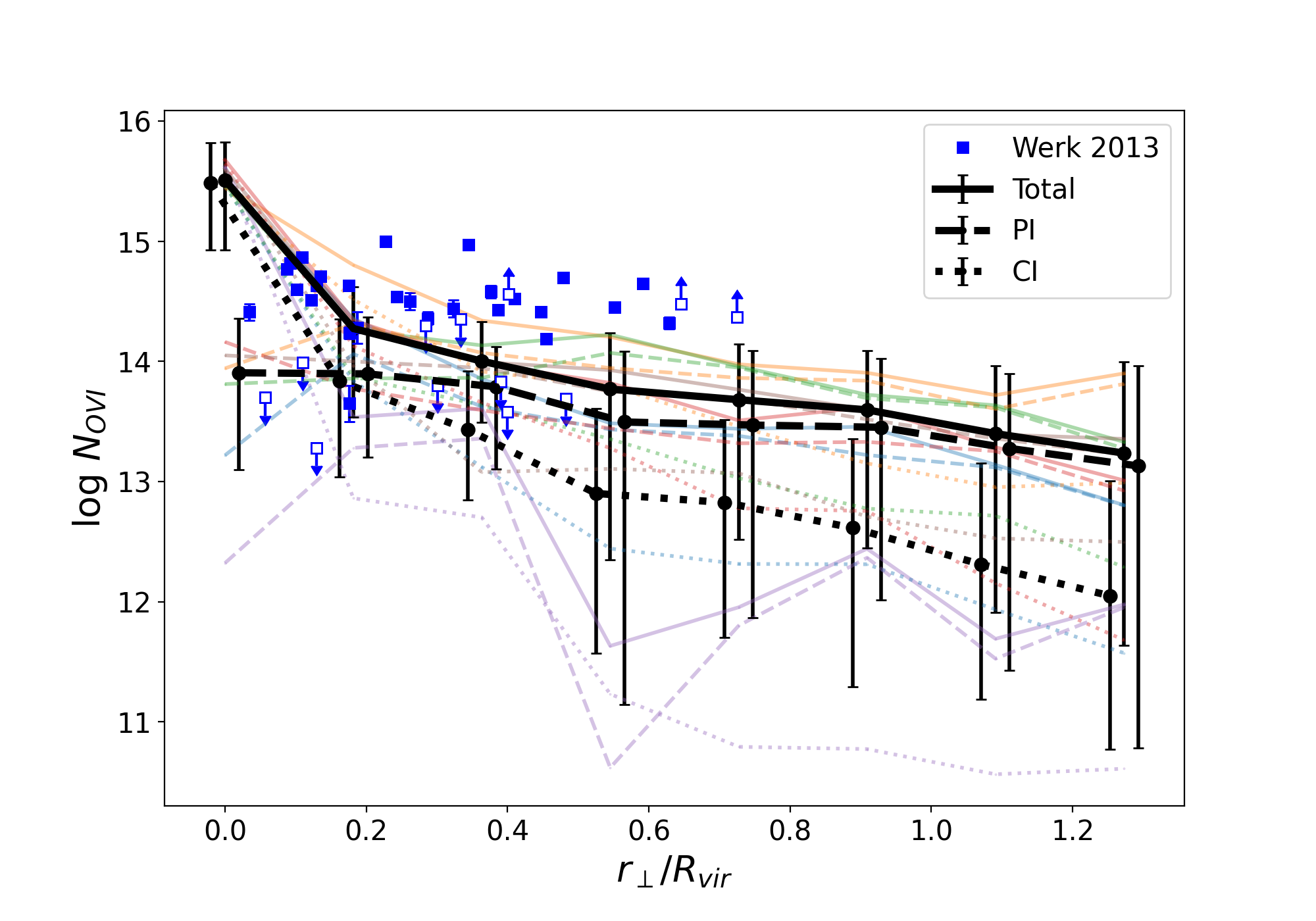}
\caption {The projection profile of several hundred sightlines per galaxy, randomly generated as described in Section \ref{sec:approach}. This is then disambiguated into total PI and CI profiles, as solid, dashed, and dotted lines, respectively. Each individual galaxy, with the same distinctive linestyles, are shown in colors. Errorbars represent the 16th and 84th percentiles, with the median value indicated by the dot.}
\label{fig:profile}
\end{figure}

We tested this in two ways. First, we used the random sightline procedure defined in Section \ref{sec:approach}. In each sightline, the total OVI column density is recorded, in addition to the fraction in the CI and PI states. The sightline can then be broken down into a `total OVI column density', a `CI OVI column density' and a `PI OVI column density'. In Figure \ref{fig:profile}, sightlines are collected by impact parameter from all six galaxies at $z=1$ and at each impact parameter, the median column density for each category is calculated. They are shown in black along with the 16th and 84th percentiles as error bars, with the total OVI in solid lines, the PI OVI in dashed lines, and the CI OVI in dotted lines (a slight offset in $x$ between the lines is included for visualization, but the data are aligned in impact parameter). The individual galaxy medians are also included in lighter colors in the background, with the same format\footnote{Colors for individual galaxies in Figure \ref{fig:profile} are the same as in Figure \ref{fig:hi-ovi}.}. Data points for OVI from \citet{werk_cos-halos_2013} are included for order-of-magnitude reference, but it is important to note that that the data are not directly comparable to the values from the VELA simulations, as the COS-Halos data is from significantly lower redshift than $z=1$, which is the lowest redshift reached by these simulations. It is worth mentioning that recent studies suggest the column density difference might also be attributable to the lack of AGN in the VELA simulation more than the further redshift evolution \citep{sanchez_not_2019}.

The main result of this study is that sightlines become dominated by PI gas at impact parameters outside $\sim 0.15 R_{\rm vir}$. While CI gas is significant inside this radius, it falls off quickly to undetectable levels in the outer halo. The CI gas predictions of $10^{12-13}$ roughly agree with the OVI columns of \citet{ji_simulations_2019}, who also found that CI OVI is found primarily in an interface layer, though unlike them we find that PI OVI is also significant in sightlines. It is also significant that the PI column density is approximately constant out to high impact parameters, only falling by approximately a factor of 2 at $r_\perp=R_{\rm vir}$, while CI column density falls by a factor of almost 1000 within the same distance.

\begin{figure}
   \centering
      \includegraphics[trim={0.65cm 0.8cm 0.7cm 1.8cm}, clip, width=0.99\linewidth]{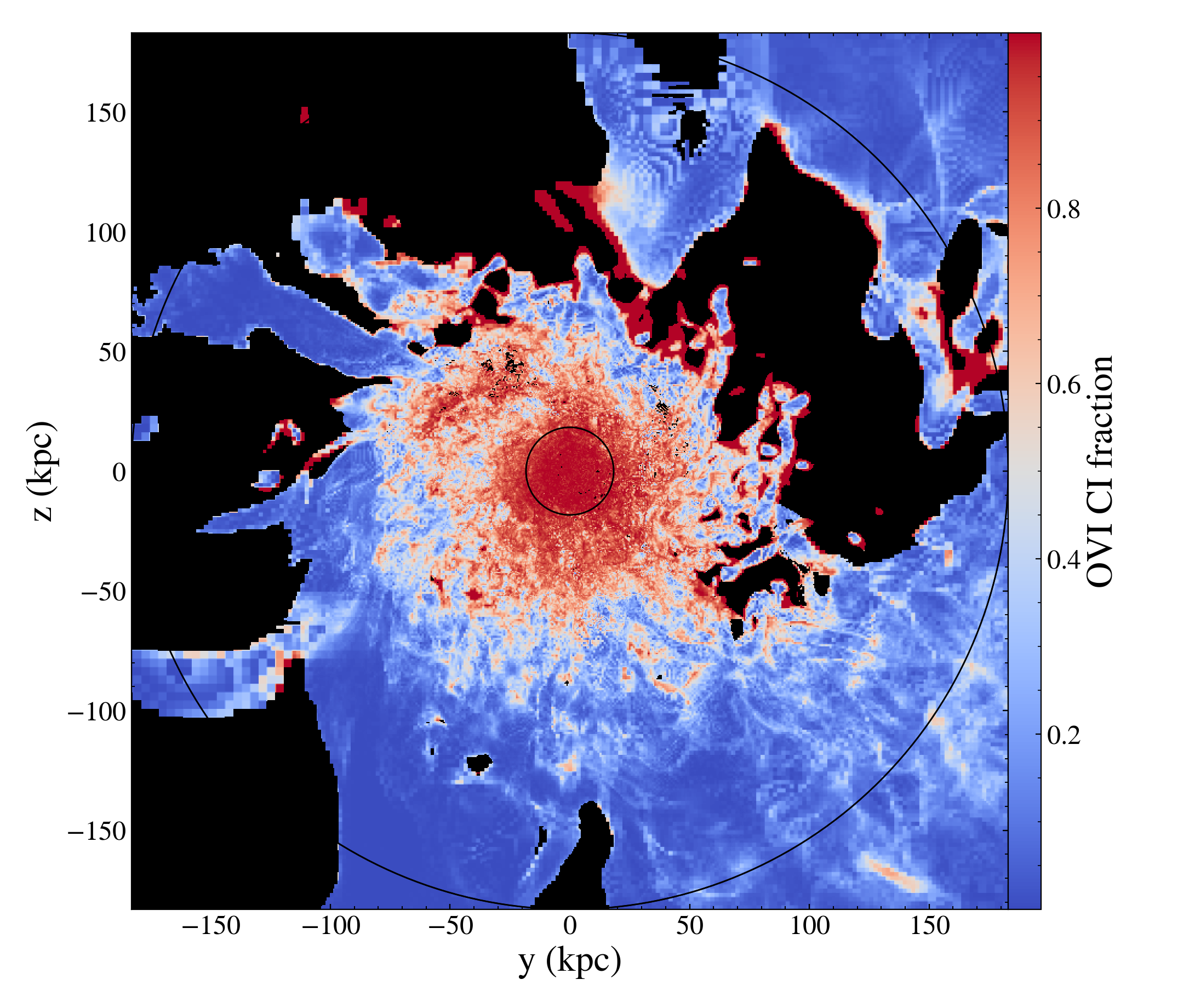}
   \caption{Fraction of gas within a projected sightline which is collisionally ionized. Each pixel represents a sightline in the $x$ direction, orthogonal to the plane of the image. Shown is VELA07 at $z=1$. A blue pixel intersects with only PI gas, a red pixel intersects only CI gas. As in Figure \ref{fig:OVIslice}, the circles represent $0.1R_{\rm vir}$ and $R_{\rm vir}$. All pixels which have a overall OVI column density $<10^{13}$ cm$^{-2}$ are blacked out, since OVI column densities $<10^{13}$ cm$^{-2}$ are not observable with COS \citet{prochaska_cos_2019}.}
   \label{fig:ci-frac}
\end{figure}

However, it is possible that the median values do not accurately convey the distribution, which due to the low filling factor of PI gas, could be very nonuniform. We also created using \textsc{yt} a projection through the full simulation volume. In Figure \ref{fig:ci-frac}, we show the CI fraction of the gas along projected sightlines. This projection has the same horizontal ($y$) and vertical ($z$) coordinates as Figure \ref{fig:OVIslice}, and the black circles continue to indicate 0.1 and 1.0 $R_{\rm vir}$. However, each pixel in this cell is the integral of all slices along the line of sight. So, a blue pixel in this image is 100 percent PI, a red pixel is 100 percent CI, and intermediate values are indicated by the color bar. We added to this image a black masking image which sets all pixels with a total OVI column density less than $10^{13}{\rm cm}^{-2}$ to black, representing nondetections. This limit is commensurate with detection limits obtained by HST/COS surveys, e.g., CASBaH \citep{prochaska_cos_2019}. If we were to adopt a threshold of 10$^{13.5}$, the picture would broadly stay the same, though slightly more of the picture would be blacked out.

We see broadly the same phenomena in this image. In the inner halo (up to about 0.2$R_{\rm vir}$), OVI is uniformly CI (red). Then, with a fairly small transitionary $r_\perp$ band (white) it switches to being nearly 100 percent PI (blue). We can see that while the detectable gas is PI outside some minimal radius, the covering fraction of all sightlines (defined here as the fraction with $N_{\rm OVI}>10^{13}$) is not 100 percent. Over all six selected VELA galaxies, the covering fraction remains $\sim 70$ per cent out to $r_\perp = R_{\rm vir}$. 

So the situation is fairly complex. The volume (at least in these relatively large galaxies which are still star-forming) is overwhelmingly dominated by CI gas, but the density of OVI within this `bulk' region is so low that it contributes almost nothing to the sightline's OVI column density. This is shown by the projection fraction (Figure \ref{fig:ci-frac}) being PI-dominated everywhere outside $0.3R_{\rm vir}$ whenever the projection isn't empty. Since we have established that a strong majority of CI gas is in fact an interface layer on PI clouds, this result is unsurprising. Wherever there is a significant amount of CI OVI, a PI gas cloud must be nearby. While these clouds may be small, their 3D structure makes them dominant over the essentially 2D surfaces of CI gas in the interface regions. Sightlines which only pass through the CI bulk region, and not through the interfaces, are visible here as the blacked out nondetections. These two images imply that almost all of the OVI which would be observed in absorption spectra at high impact parameters is PI. 

The inner halo (0.1-0.3$R_{\rm vir}$) was found to be highly irregular and different from the outer halo in cosmological simulations similar to VELA in \citet{danovich_four_2015}. It is also possible that in this inner halo region, the fixed size (1 or 2 kpc) of the interface might be larger than strictly necessary, and could sample gas which is not dynamically connected to the PI gas it happens to be near and we will need to be cautious interpreting the results in light of the interface layer containing most OVI. Within this region, there is significant warm-hot, metal-rich gas outflowing due to stellar feedback, and its effects on the overall dynamics of the gas distribution in Table \ref{tab:2} are substantial. All these effects noted, sightlines in the inner halo were found to be almost entirely CI, and we will examine the reason for this transition in more detail in Section \ref{sec:other_ions}.

\begin{figure*}
\includegraphics[trim={2.5cm 0.6cm 3.5cm 2.1cm}, clip, width=0.99\linewidth]{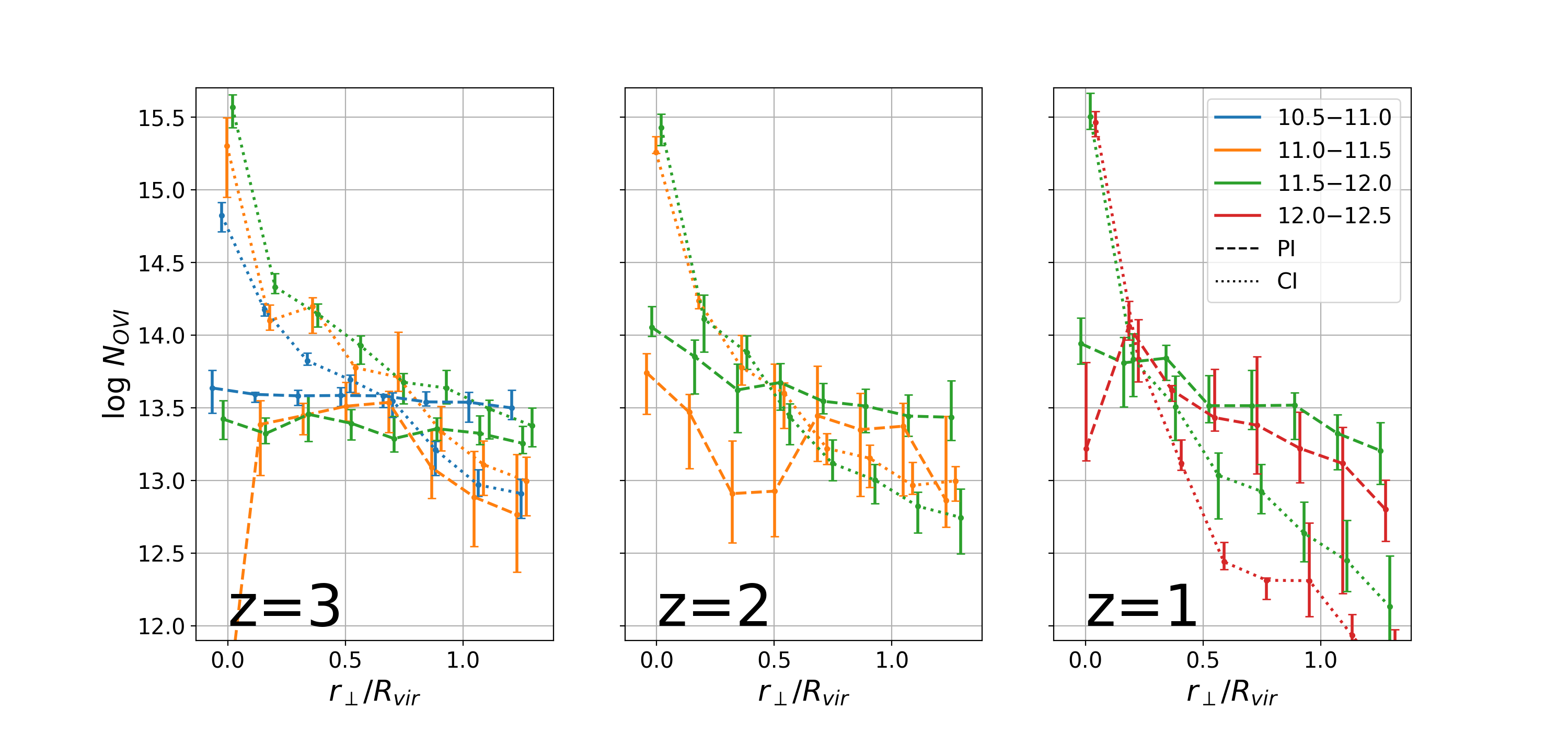}
\caption {Evolution of OVI column densities with redshift and mass. Mass bins indicated in legend are in units of log M$_\odot$. The left panel is a snapshot of all galaxies at redshift 3, the center panel is the same at redshift 2, and the right is the same at redshift 1. Errorbars represent $\pm$ 10 percentiles.}
\label{fig:massdep}
\end{figure*}

\subsection{Halo mass and redshift dependence}\label{sec:mass-redshift-dependence}

Now we will compare how the effects shown in Figure \ref{fig:profile} change with mass and redshift. In \citetalias{roca-fabrega_cgm_2019} the mass and redshift dependence of the ionization mechanism of OVI in the CGM was as follows (see \citetalias{roca-fabrega_cgm_2019}, Figure 14). All galaxies start out with their OVI population entirely CI-dominated. This is a function of three effects: First, the low ionizing background at high ($z>2.5$) redshift, second, the fact that at high redshift, the cold inflows are almost metal-free, and third, at higher redshift the streams are denser and more self-shielded \citep{mandelker_cold_2018,mandelker_lyalpha_2020}.  The galaxies then experience a decrease in their CI fraction with time as the ionizing background becomes more significant and streams become less dense. As galaxies approach redshift zero, their OVI ionization mechanisms diverge according to their mass. Low-mass galaxies end up completely PI at late times, while high mass galaxies become mostly CI again, following the formation of a virial shock which heats up most of the CGM.

We can see some of the same effects in the time-series sightline projections (Figure \ref{fig:massdep}). Here we repeat the procedure of Figure \ref{fig:profile}, including showing the profile of the median sightline with error bars representing the 40th and 60th percentiles, except binning the galaxies into mass bins of 0.5 dex at specific snapshots instead of combining all sightlines together into a single `overall' curve. The smaller errorbars here compared to Figure \ref{fig:profile} are to avoid overlapping lines. All data points are offset slightly in $r_\perp$ so the error bars are visible, but should be read as vertically aligned in the apparent groups. The substantial bias of impact parameter profiles towards outer-CGM gas, as discussed in Section \ref{sec:sightlines}, means PI gas still dominates for most redshifts and masses. However, the trend from \citetalias{roca-fabrega_cgm_2019} is evident in the form of the decreasing `transition impact parameter' where PI gas becomes dominant. At $z=3$, we see that only the smallest galaxies (blue line) have such a transition, and the larger galaxies (orange and green lines) remain CI-dominated out to $r_\perp = R_{\rm vir}$. Moving on to redshift 2, we see that both of the available mass bins have roughly the same crossing point at $\sim 0.6 R_{\rm vir}$, and CI dominates inside that impact parameter, PI dominates outside. At low redshift ($z=1$), we see that this transition from CI to PI-dominated sightlines happens at $\sim 0.2 R_{\rm vir}$ as shown before. It is also worth noting that the CI gas seems to drop much more dramatically with impact parameter at redshift 1, and generally the OVI column density drops below 10$^{13.5}$ at the outer halo for the first time. It is worth mentioning that as the galaxies evolve with time, the virial radii are increasing, so the decline in CI gas in the outer halo might be partially explained as new PI OVI being enclosed in $\Rv$ without any accompanying CI OVI.

We see that there is not a significant mass dependence of either the CI or the PI gas. Unlike in \citetalias{roca-fabrega_cgm_2019}, at $z=3$ they diverge mostly for sightlines which pass through or close to the galaxy, and at z=1 and z=2 there is little change in column density with mass between the available bins. 

In this set of simulations, we are not studying any galaxies which have a smaller virial mass than $M_{\rm vir}=10^{11}$M$_\odot$. As presented in \citetalias{roca-fabrega_cgm_2019}, these small galaxies will allow inflows to reach all the way to the disc, which means that this model would suggest they are all PI. Studying whether this is generally true in smaller galaxies will be the subject of future work.

\subsection{Comparison with Observations}\label{sec:observations}
Using a phenomenological analysis of the COS-Halos data, \citetalias{stern_universal_2016} proposed that cool and relatively low density clouds produce the observed OVI columns of $\lesssim10^{14.5} {\rm cm}^{-2}$ and a comparable amount of neutral hydrogen ($N_{\rm HI}/N_{\rm OVI}\sim3$), while higher density clouds embedded in or at smaller scales than the OVI clouds produce low ions and larger HI columns. This density structure suggests that sightlines with $N_{\rm HI}\lesssim10^{15} {\rm cm}^{-2}$ intersect the OVI clouds but not the low-ion clouds, and hence $N_{\rm HI}$ and $N_{\rm OVI}$ should be correlated in these sightlines, while sightlines with $N_{\rm HI}\gg10^{15} {\rm cm}^{-2}$ intersect both the OVI clouds and the low-ion clouds, and hence $N_{\rm OVI}$ should be independent of $N_{\rm HI}$. In Figure 10 we show that sightlines through the VELA simulations follow the same pattern. This indicates that the 'global' version of the \citetalias{stern_universal_2016} model, where OVI and $N_{\rm HI}\sim10^{15}$ columns originate from the outer halo and low-ions and $N_{\rm HI}\gg10^{15}$ columns originate from the inner halo, replicates the behavior in VELA.

\begin{figure}
\includegraphics[trim={1.2cm 0.5cm 1.9cm 2.0cm}, clip, width=0.99\linewidth]{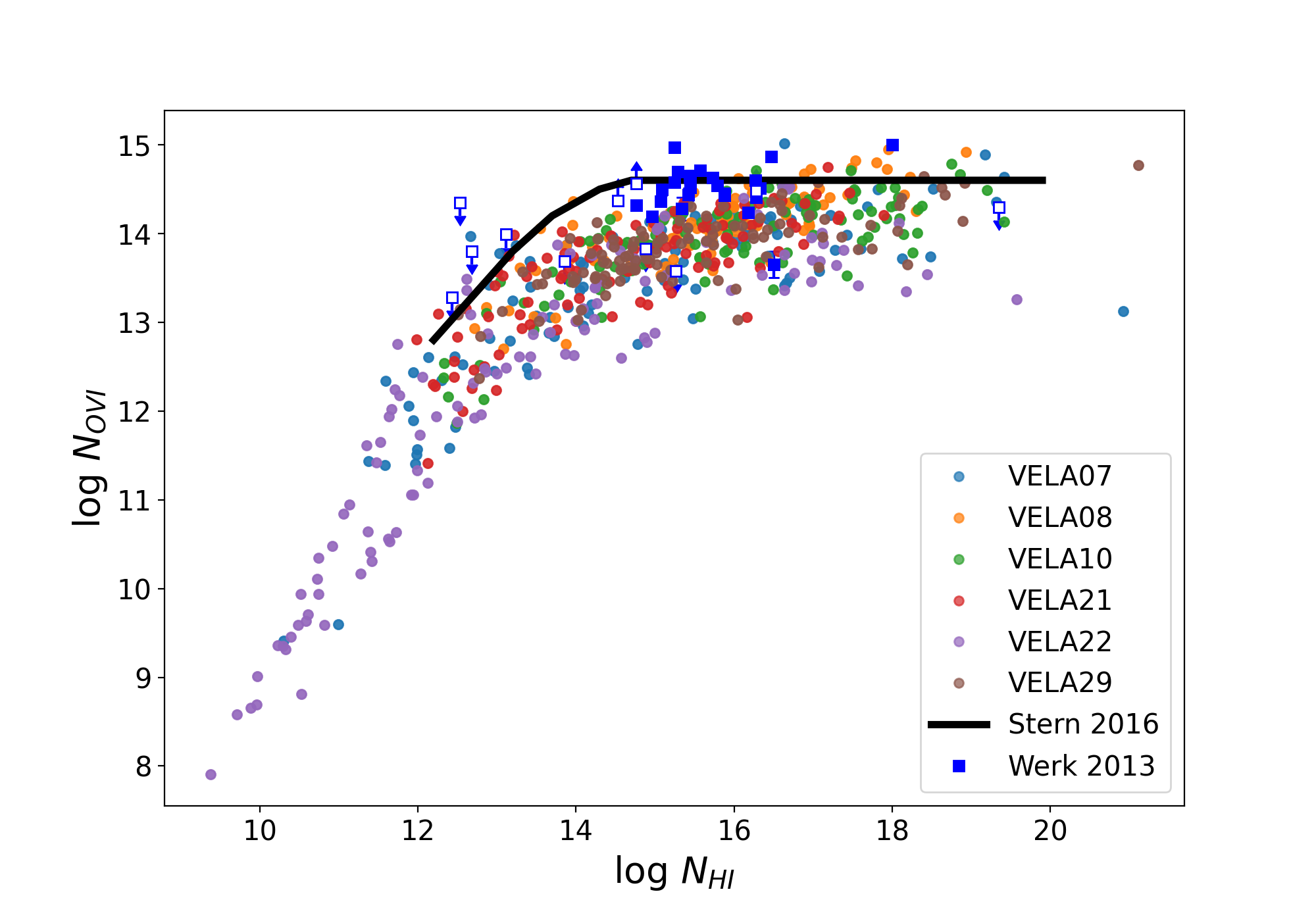}
\caption {Comparison of the HI vs OVI column densities of all of the sightlines through the different VELA simulations. The black curve shows the theoretical prediction from the phenomenological model presented in \citetalias{stern_universal_2016}, fitted to the COS-halos data (\citet{werk_cos-halos_2013}, blue).}
\label{fig:hi-ovi}
\end{figure}

It should be noted that simulated HI distributions in the CGM are resolution-dependent \citep{hummels_impact_2019}, and so it is possible that $N_{\rm HI}$ is not converged. This however should only have an effect at the highest $N_{\rm HI}$, where the HI-OVI curve shown in Figure \ref{fig:hi-ovi} has already flattened out. Also, we find that $N_{\rm OVI}$ in the VELA simulations is a factor of $\sim 3$ lower than in COS-Halos, potentially due to the higher redshift of $z\sim1$ analyzed in VELA relative to $z\sim0.2$ in COS-Halos.

\begin{figure}
\includegraphics[trim={1.1cm 0.5cm 1.9cm 1.9cm}, clip, width=0.99\linewidth]{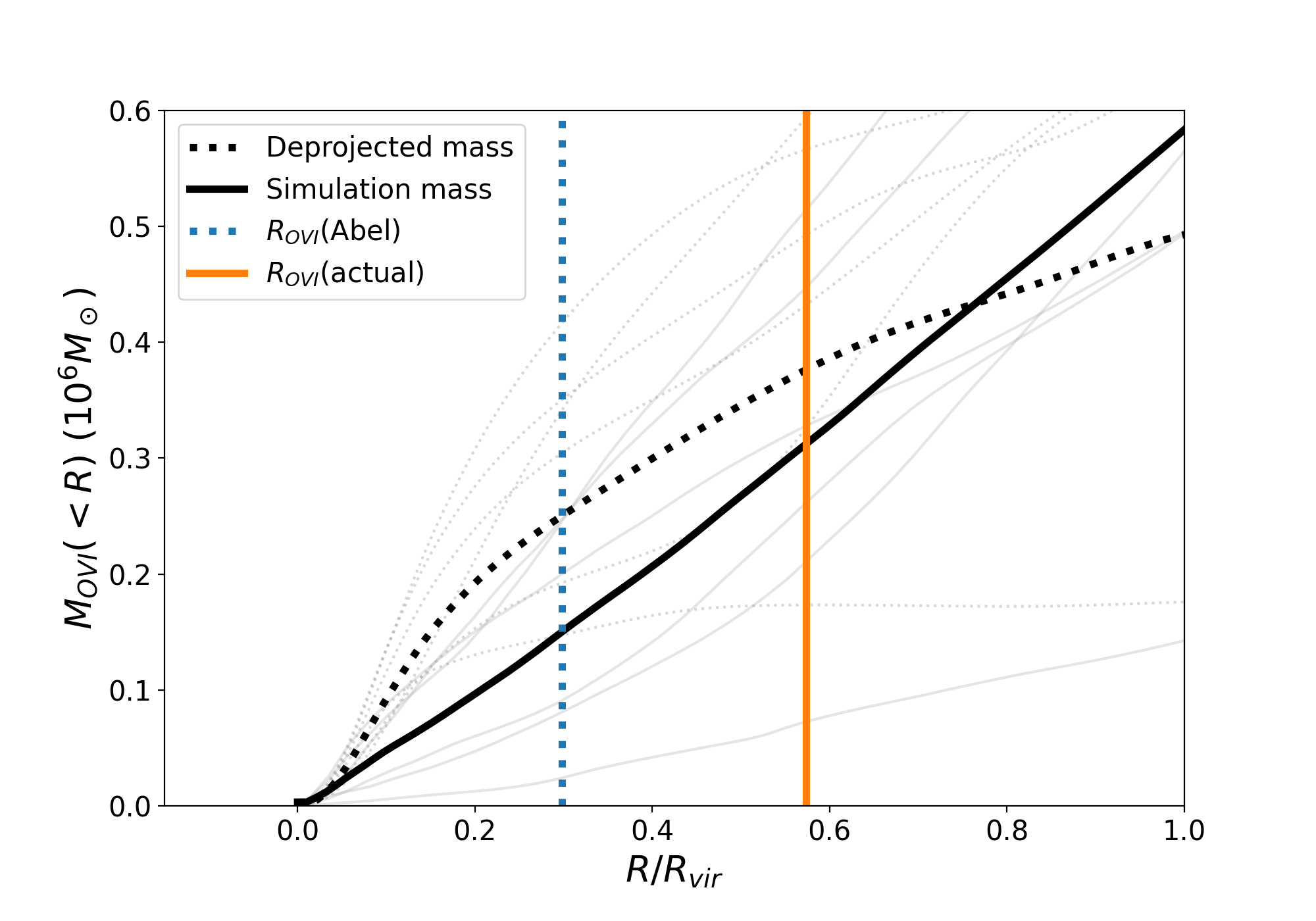}
\caption {Using the data from Figure \ref{fig:profile}, an inverse Abel transformation is performed on the mock sightlines through the stacked VELA simulations to determine an approximate mass of OVI within the CGM, and a half-mass radius $R_{\rm OVI}$ (dotted lines). This is compared to the actual distribution and half-mass radius from integrating over the simulation directly (solid lines).}
\label{fig:Abel}
\end{figure}

We can also check whether the sightlines allow us to infer correctly the 3D distribution of OVI. \citetalias{stern_does_2018} showed that the column densities observed from COS could, under an assumption of spherical symmetry, be used to extrapolate the total OVI mass in the CGM as a cumulative function of radius. Assuming that all galactic CGMs from COS-Halos were broadly similar, one can use an inverse-Abel transformation on the column densities of OVI to predict the total mass of OVI in the CGM of an average galaxy, relative to $R{\rm vir}$ \citep{mathews_circumgalactic_2017,stern_does_2018}. In \citetalias{stern_does_2018}, the purpose of this was to make the argument that the median OVI ion's radius, or $R_{\rm OVI}$, actually exists outside half of the virial radius, and so is more emblematic of the outer CGM than the inner part, even in sightlines with impact parameters less than $R_{\rm OVI}$.  This makes the assumption, however, that the CGM is spherically symmetric. In Figure \ref{fig:profile}, we showed the median OVI column densities for a set of mock sightlines within the simulation. We will assume those median results are spherical and then apply the same inverse-Abel transformation algorithm to them as in \citetalias{stern_does_2018}. This can be compared to the real distribution of OVI gas. We find in Figure \ref{fig:Abel} that the inverse-Abel transform indeed approximately recovers the actual mass of OVI within the virial radius to within $\sim$20 per cent. An interesting distinction between the two curves is rather their shape. We find that the deprojected curve appears to be concave down, so it would be overrepresented in the inner CGM and underrepresented in the outer CGM, while the real OVI gas distribution is approximately linear out to the virial radius. Its different concavity (compare Figure \ref{fig:Abel}, with \citetalias{stern_does_2018}, Figure 1, top) leads to our placing the (deprojected) $R_{\rm OVI}$ closer to the inner CGM than the actual $R_{\rm OVI}$. This suggests then that in \citetalias{stern_does_2018} itself, it is likely that the prediction for $R_{\rm OVI}$ was an underestimate, because their deprojection was indeed concave down and we have shown that a linear radial profile in real space will lead to a concave-down profile in deprojection-space. This means that most of the OVI in real observed sightlines might be near the edge of the virial radius, and there may even be a significant component in the IGM, if the virial radius is taken to be the boundary of the CGM.

\begin{figure}
\includegraphics[width=\linewidth]{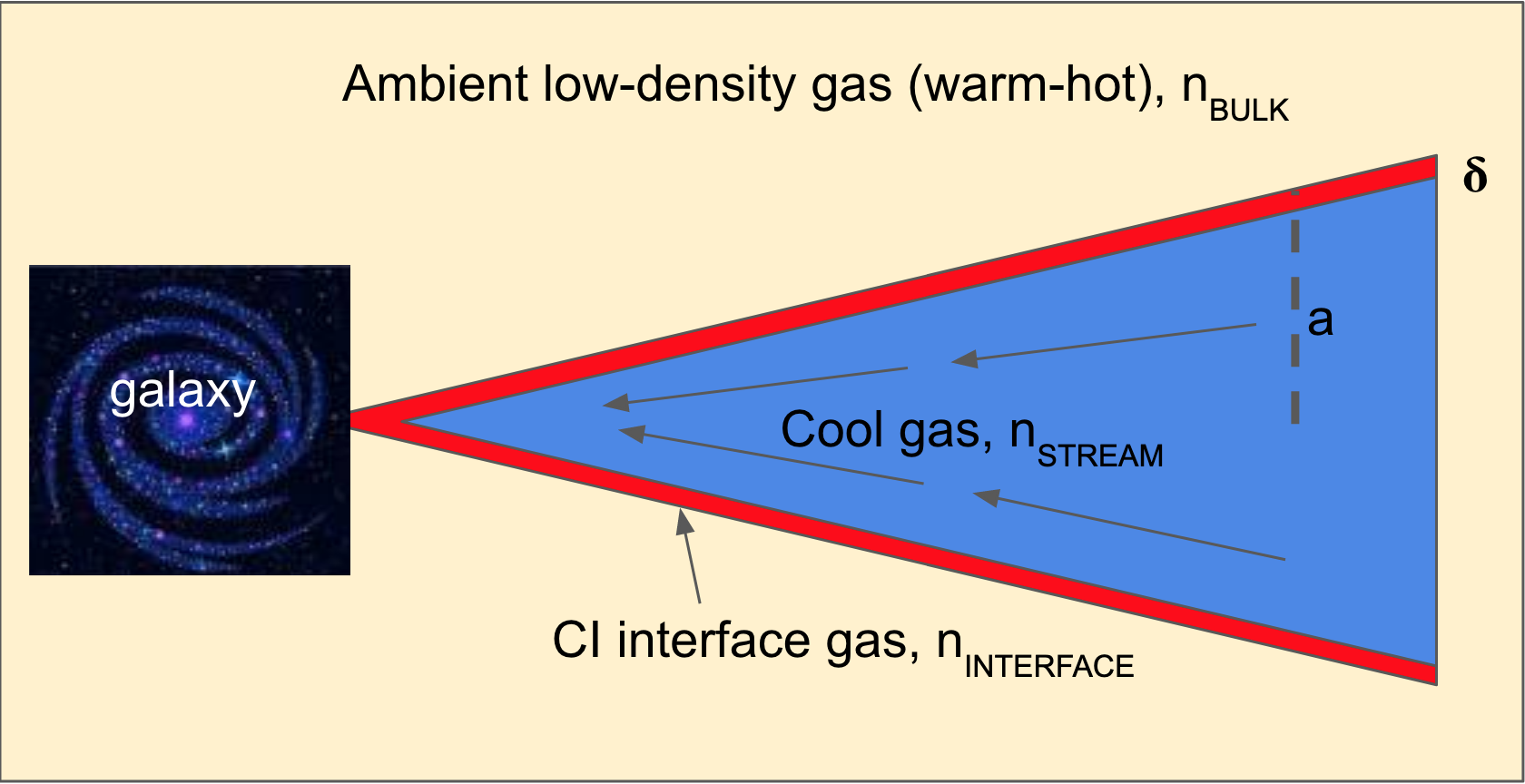}
\caption {The model suggested by the inflow patterns for the CGM. While it seems that there is a coincidence between the inflows and the PI-CI cutoff for OVI, other ions, especially CIV, also appear to be regulated by these structures as in Figure \ref{fig:otherions}.}
\label{fig:model}
\end{figure}

In addition, our results that OVI traces cool inflows, combined with the \citet{tumlinson_large_2011} result that OVI is absent around quenched galaxies (albeit at lower redshift), may be evidence that the feedback mechanism which quenches galaxies also directly affects the cool inflows. We plan to study this effect in future work, using simulations which reach lower redshifts and higher masses.


\section{Physical Interpretation of the Interface Layer}\label{sec:model_2}

There is existing literature regarding how the structure of galaxy formation is strongly regulated by inflows from the cosmic web into the galaxy through the CGM \citep[e.g.][and references therein]{keres_how_2005,dekel_galaxy_2006,dekel_cold_2009,fox_gas_2017}. We suggest that the metal distribution of the CGM might be governed by the same structures, and propose a three-phase model for OVI and other ions in the CGM. A cartoon picture is shown in Figure \ref{fig:model}. In this model, there are three regions of the CGM: the inside of the cool-inflow cones (hereafter the cold component, or cold streams), the outside (hereafter the hot component, or hot CGM), and the interface between these two components. The interaction between the inflowing cold streams and the ambient hot CGM induces Kelvin-Helmholtz instabilities (KHI) and thermal instabilities at the interface, causing hot gas to become entrained in the flow through a strongly cooling turbulent mixing layer of intermediate densities and temperatures (\citealp{mandelker_instability_2020}, hereafter \citetalias{mandelker_instability_2020}). We posit that the CI interface layer we find in our simulations represents precisely such a mixing layer. 
The general properties of radiatively cooling interface layers induced by shear flows were studied in \citet{ji_simulations_2019} and \citet{fielding_multiphase_2020}. The conditions of the cold streams and hot CGM, which set the boundary conditions for the interface region, as a function of halo mass, redshift, and position within the halo were studied in \citet{mandelker_lyalpha_2020} (hereafter \citetalias{mandelker_lyalpha_2020}), based on \citetalias{mandelker_instability_2020}. In this section, we combine the insights of these studies to explain the physical origin and the properties of the multiphase structure seen in our simulations. We begin in \se{streams_theory} by summarizing our current theoretical understanding of the evolution of cold streams in the CGM of massive high-$z$ galaxies, as they interact with the ambient hot gaseous halo. In \se{streams_sims} we examine the properties of the different CGM phases identified in our simulations, in light of this theoretical framework. Finally, in \se{other_ions} we use these insights to model the distribution of OVI and other ions in the CGM of massive $z\sim 1$ galaxies.

\subsection{Theoretical Framework}
\label{sec:streams_theory}

\subsubsection{KHI in Radiatively Cooling Streams}

Using analytical models and high-resolution idealized simulations, these have focused on pure hydrodynamics in the linear regime \citep{M16} and the non-linear regime in two dimensions \citep{P18,Vossberg19} and three dimesnsions \citep{M19}. Others have incorporated self-gravity \citep{Aung19}, idealized MHD \citep{Berlok19}, radiative cooling \citepalias{mandelker_instability_2020}, and the gravitational potential of the host dark matter halo \citepalias{mandelker_lyalpha_2020}. We begin by summarizing the main findings of  \citetalias{mandelker_instability_2020} regarding KHI in radiatively cooling streams. There, we considered a cylindrical stream\footnote{The relation between cylindrical streams and the conical nature of \fig{model} is addressed in \se{stream_halo}.} with radius $\Rs$, density $\rhos$, and temperature $\Ts$, flowing with velocity $V_{\rm s}$ through a static background ($V_{\rm b}=0$) with density $\rhob$ and temperature $\Tb$. The stream and the background are assumed to be in pressure equilibrium, so $\chi \equiv \rhos/\rhob = \Tb/\Ts$, where we have neglected differences in the mean molecular weight in the stream and the background. The Mach number of the flow with respect to the sound speed $\cb$ in the background is $\Mb\equiv V_{\rm s}/\cb$.

The shear between the stream and the background induces KHI, which leads to a turbulent mixing region forming at the stream-background interface. The characteristic density and temperature in this region are \citep{begelman_turbulent_1990,gronke_growth_2018} 
\be 
\label{eq:nmix}
\rho_{\rm mix}\sim \left(\rhob \rhos\right)^{1/2} = \chi^{-1/2}\rhos,
\ee 
\be 
\label{eq:Tmix}
T_{\rm mix}\sim \left(\Tb \Ts \right)^{1/2} = \chi^{1/2}\Ts.
\ee

In the absence of radiative cooling, the shear region engulfs the entire stream in a timescale
\be 
\label{eq:tshear}
t_{\rm shear} = \frac{\Rs}{\alpha V_{\rm s}},
\ee
{\no}where $\alpha \sim (0.05-0.1)$ is a dimensionless parameter that depends on the ratio of stream velocity to the sum of sound speeds in the stream and background, $M_{\rm tot}=\Vs/(\cs+\cb)$ \citep{P18,M19}. When radiative cooling is considered, the non-linear evolution is determined by the ratio of $t_{\rm shear}$ to the cooling time in the mixing region, 
\be 
\label{eq:tcool_mix}
\tcm = \frac{k_{\rm B}T_{\rm mix}}{(\gamma-1)\,n_{\rm mix}\,\Lambda(T_{\rm mix})},
\ee
{\no}with $\gamma=5/3$ the adiabatic index of the gas, $k_{\rm B}$ Boltzmann's constant, $n_{\rm mix}$ the particle number density in the mixing region, and $\Lambda(T_{\rm mix})$ the cooling function evaluated at $T_{\rm mix}$. If $t_{\rm shear}<\tcm$, then KHI proceeds similarly to the non-radiative case, shredding the stream on a timescale of $t_{\rm shear}$ \citep{M19}. However, if $\tcm<t_{\rm shear}$, hot gas in the mixing region cools, condenses, and becomes entrained in the stream \citepalias{mandelker_instability_2020}. In this case, KHI does not destroy the stream. Rather, it remains cold, dense and collimated until it reaches the central galaxy. Similar behaviour is found in studies of spherical clouds \citep{gronke_growth_2018,gronke_how_2020,Li_Hopkins_20}, and planar shear layers \citep{ji_simulations_2019,fielding_multiphase_2020}.

Streams with $\tcm<t_{\rm shear}$ grow in mass by entraining gas from the hot CGM as they travel towards the central galaxy. The stream mass-per-unit-length (hereafter line-mass) as a function of time is well approximated by \citepalias{mandelker_instability_2020} 
\be 
\label{eq:mass_rad_approx}
m(t) = m_0\left(1+\dfrac{t}{t_{\rm ent}}\right),
\ee
{\no}where $m_0=\pi\Rs^2\rhos$ is the initial stream line-mass, and the entrainment timescale is 
\be 
\label{eq:tent}
t_{\rm ent} = \frac{\chi}{2}\left(\frac{t_{\rm cool}}{\tsc}\right)^{1/4} \tsc, 
\ee 
{\no}with $\tsc=2\Rs/c_s$ the stream sound crossing time, and $t_{\rm cool}$ the minimal cooling time of material in the mixing layer. which in practice has a distribution of densities and temperatures rather than being a single phase described by \equs{nmix}-\equm{Tmix}. If the stream is initially in thermal equilibrium with the UV background, the minimal cooling time occurs approximately at $T=1.5\Ts$, but any temperature in the range $\sim (1.2-2)\Ts$ works equally well \citepalias{mandelker_instability_2020}. The density is given by assuming pressure equilibrium. This mass entrainment causes the stream to decelerate, due to momentum conservation. A large fraction of the kinetic and thermal energy dissipated by the stream-CGM interaction is emitted in Ly$\alpha$, which may explain the extended Ly$\alpha$ blobs observed around massive high-$z$ galaxies (\citealp{Goerdt10}, \citetalias{mandelker_lyalpha_2020}). 

\begin{figure*}
\includegraphics[trim={3.7cm 0.5cm 4.0cm 0cm}, clip, width=0.99\textwidth]{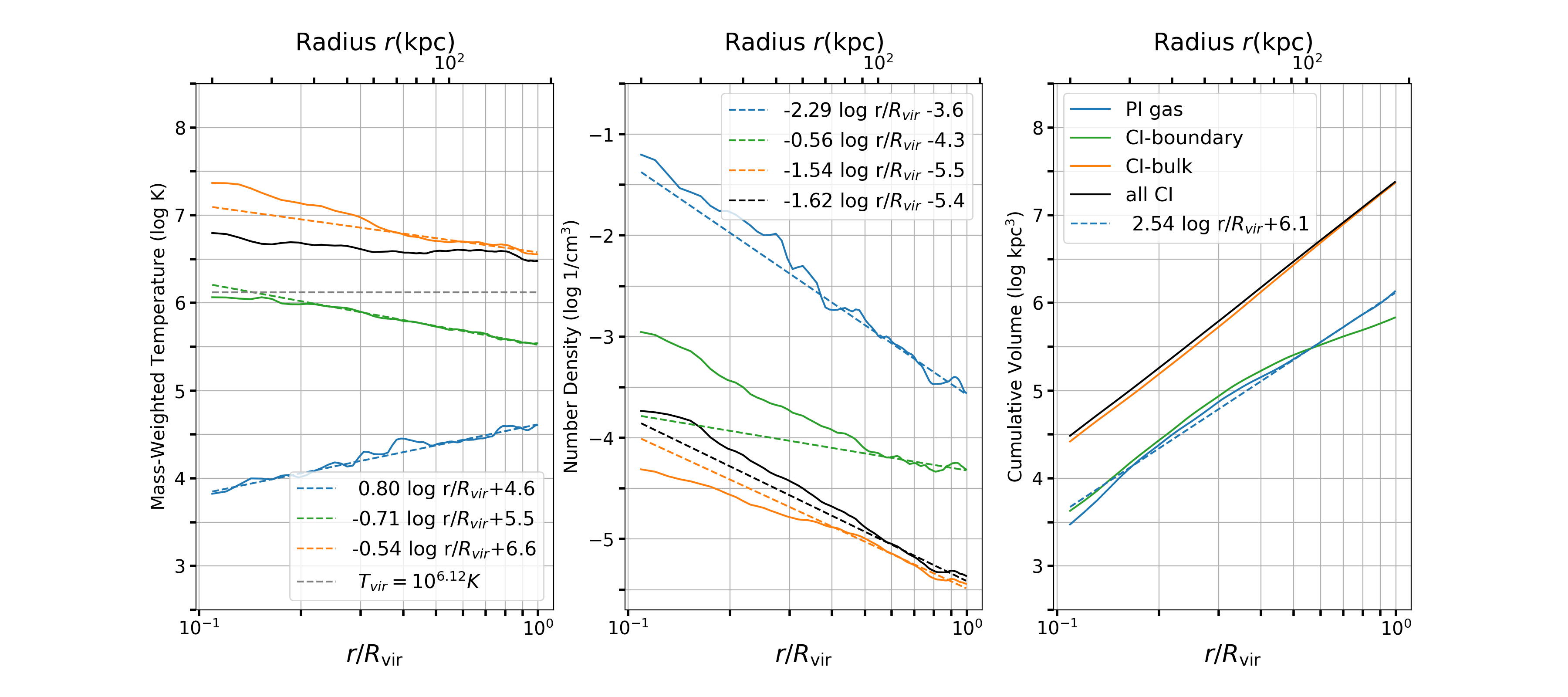}
\caption{Radial profiles of physical properties in the three OVI phases in the CGM of VELA07 at $z=1$. Blue lines represent PI gas, associated with cold streams, orange line represent bulk CI gas, associated with the hot CGM, and green lines represent CI interface gas, associated with the mixing layer between the cold streams and hot CGM. Dotted lines represent best-fit power law relations for the radial profile of the same color and type, fit in the radial range $r=(0.5-1)\Rv$, and the fits themselves are listed in the panel legends. We also show profiles for all CI gas in black. \textit{Left:} Temperature profiles, showing the mass-weighted average temperature in each radial bin. 
\textit{Centre:} Gas density profiles within each radial bin. 
\textit{Right:} Cumulative volume occupied by each phase at radii $\le r$. 
Only the stream volume power law is shown.}
\label{fig:streams_profiles_1}
\end{figure*}

\subsubsection{Stream Evolution in Dark Matter Halos}
\label{sec:stream_halo}

In order to address the evolution of streams in dark matter halos, \citetalias{mandelker_lyalpha_2020}, following earlier attempts \citep{dekel_galaxy_2006,dekel_cold_2009}, developed an analytical model for the properties of streams as a function of halo mass and redshift. We here focus on $\sim 10^{12}\msun$ halos at $z\sim 1$ (\tab{1}), and refer readers to \citetalias{mandelker_lyalpha_2020} for more general expressions. Near the halo edge, at $\Rv$, the streams are assumed to be in approximate thermal equilibrium with the UV background, yielding temperatures of $T_{\rm cold} \sim 2\times 10^4{\rm K}$. The temperature in the hot CGM is assumed to be of order the virial temperature, 
\be 
\label{eq:Tvir}
T_{\rm hot} \sim \Tv\simeq 10^6\K~M_{12}^{2/3}(1+z)_2,
\ee
{\no}with $M_{12}=\Mv/10^{12}\Msun$ and $(1+z)_2=(1+z)/2$. The stream and the hot CGM are assumed to be in approximate hydrostatic equilibrium. Accounting for order-unity uncertainties in the above quantities, the density contrast between the stream and the hot CGM is predicted to be in the range $\chi\sim (20-200)$, with a typical value of $\sim 70$. 

The density of the hot gas is constrained by the dark matter halo density in the halo outskirts, the Universal baryon fraction, and the fraction of baryonic matter in the hot CGM component, which has constraints from observations and cosmological simulations. Together with the $\chi$ values quoted above, this gives the density in streams as they enter $\Rv$. This is predicted to be $n_{\rm s} \sim (3\times 10^{-4}-0.01)\cmc$, with a typical value of $10^{-3}\cmc$.

In \citetalias{mandelker_lyalpha_2020}, the stream is assumed to enter the halo on a radial orbit, with a velocity comparable to the virial velocity, 
\be 
\label{eq:Vvir}
\Vv\simeq 163\kms~M_{12}^{1/3}\,(1+z)_2^{1/2}.
\ee
The mass flux entering the halo along the stream is given by the total baryonic mass flux entering the halo, and the fraction of this mass flux found along streams, where one dominant stream typically carries $\sim$half the inflow, while three streams carry $\gsim 90\%$ \citep{Danovich_2012}. The stream density, velocity, and mass flux can together be used to constrain the stream radius. This is predicted to be $\Rs/\Rv \sim (0.03-0.5)$, with a typical value of $\sim 0.2$, and where the virial radius is given by 
\be 
\label{eq:Rvir}
\Rv\simeq 150\kpc~M_{12}^{1/3}\,(1+z)_2^{-1}.
\ee
{\no}Inserting the above constraints for the stream and hot CGM properties into \equs{nmix}-\equm{tcool_mix} leads to the conclusion that $\tcm<t_{\rm shear}$ in virtually all cases, even if the streams are nearly metal-free \citepalias{mandelker_lyalpha_2020}. Streams are thus expected to survive until they reach the central galaxy, and grow in mass along the way.

Within the halo, at $0.1<r/\Rv<1$, \citetalias{mandelker_lyalpha_2020} assumed both the stream and the background to be isothermal, and to have a density profile described by a power law, 
\be
\label{eq:rho_prof}
\rho\propto x^{-\beta}, 
\ee
{\no}with $x\equiv r/\Rv$ and $1<\beta<3$. The stream and halo thus maintain pressure equilibrium at each halocentric radius, with a constant density contrast $\chi$. The stream is assumed to be flowing towards the halo centre, growing narrower along the way. The stream radius at halocentric radius $r$ is 
\be 
\label{eq:rs_prof}
a = \Rs \left(\frac{m(r)}{m_0}\right)^{1/2}x^{\beta/2},
\ee 
{\no}with $m(r)$ the stream line-mass at halocentric radius $r$, $m_0$ the line-mass at $\Rv$, and $\Rs$ the stream radius at $\Rv$. In general, $m(r)>m_0$ due to the mass entrainment discussed above. However, in practice, the line mass of streams on radial orbits in $10^{12}\msun$ halos at $z\sim 1$ grows by only $\sim (5-40)$ percent by the time the stream reaches $0.1\Rv$ \citepalias{mandelker_lyalpha_2020}. We can thus approximate $a\propto x^{\beta/2}$. In this case, it is straightforward to show that the cumulative volume occupied by the stream interior to a halocentric radius $r$ is 
\be 
\label{eq:Vol_s}
{\rm Vol_s}(r) = \frac{\pi \Rv^3}{\beta+1}\left(\dfrac{\Rs}{\Rv}\right)^2\left(\dfrac{r}{\Rv}\right)^{\beta+1}.
\ee 

\citetalias{mandelker_lyalpha_2020} assumed that the mass entrainment rates derived by \citetalias{mandelker_instability_2020} (\equsnp{mass_rad_approx}-\equmnp{tent}) could be applied locally at each halocentric radius. When doing so, they used the scaling $t_{\rm cool}\propto n^{-1}\propto x^{\beta}$ and $\tsc\propto a \propto x^{\beta/2}\mu^{1/2}$, with $\mu\equiv m(r)/m_0$. They then derived equations of motion for the stream within the halo, where the deceleration induced by mass entrainment counteracts the acceleration due to the halo potential well. These equations were solved simultaneously for the radial velocity and the line-mass of streams as a function of halocentric radius. For $10^{12}\msun$ halos at $z\sim 1$, the line-mass at $0.1\Rv$ was found to be $\sim (5-40)$ percent larger than at $\Rv$, while the radial velocity was $\sim (75-98)$ percent of the free-fall velocity.

\begin{figure*}
\includegraphics[trim={3.0cm 0.85cm 4.0cm 0cm}, clip, width=0.99\textwidth]{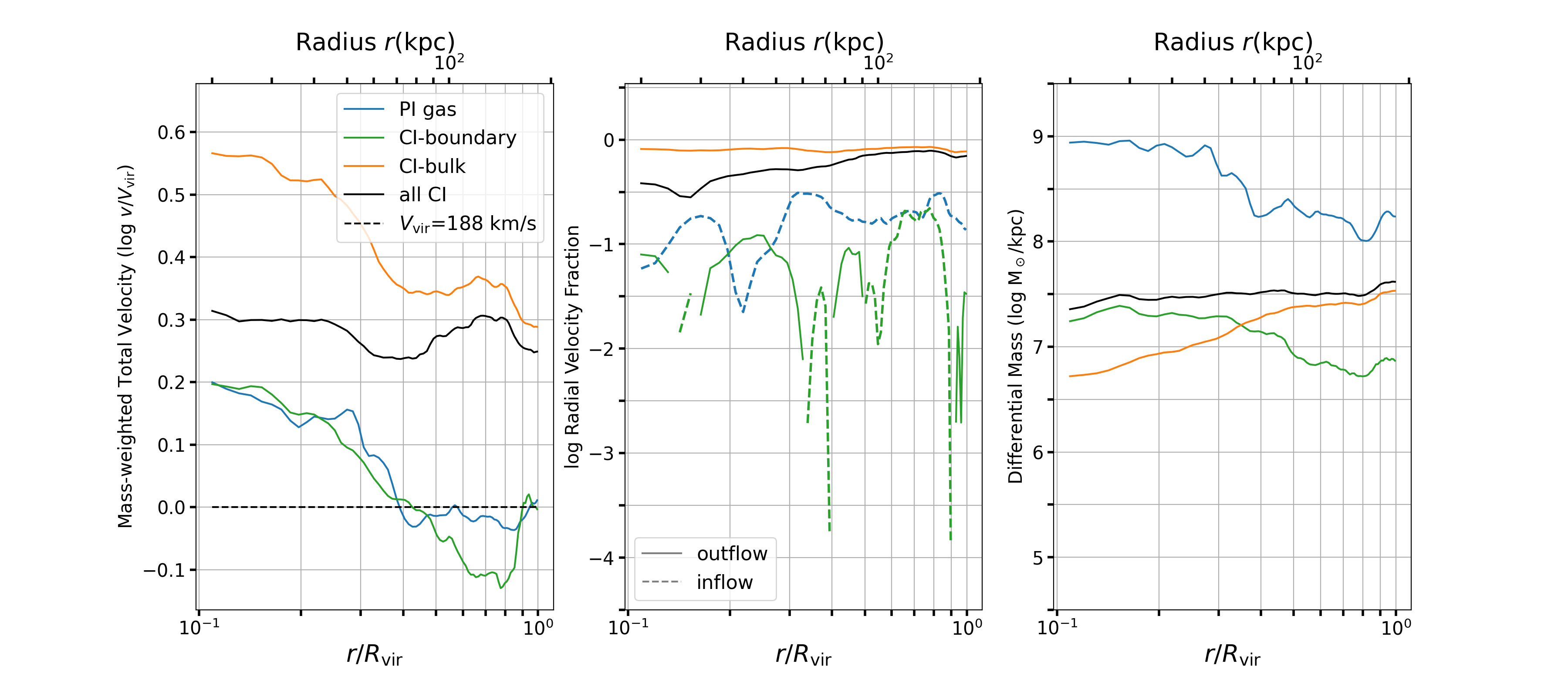}
\caption{Radial profiles of dynamical properties in the three OVI phases in the CGM of VELA07 at $z=1$. Line colours are as in \fig{streams_profiles_1}. \textit{Left:} Total velocity in each component, normalized by the halo virial velocity, $\Vv$. 
\textit{Centre:} Radial velocity, normalized to total velocity in each radial bin. Solid (dashed) lines represent outflowing (inflowing) gas. 
\textit{Right:} Mass-per-unit-length (line-mass) within each state. 
}
\label{fig:streams_profiles_2}
\end{figure*}

\subsubsection{Turbulent Mixing Layer Thickness}

Several recent studies have examined the detailed physics behind the growth of turbulent mixing layers and the flux of mass, momentum, and energy through them \citep{P18,ji_simulations_2019,fielding_multiphase_2020}. Using idealized numerical simulations and analytical modeling, these works considered a simple planar shear layer between two semi-infinite domains, without \citep{P18} and with \citep{ji_simulations_2019,fielding_multiphase_2020} radiative cooling. While this is different than the cylindrical geometry we have thus far considered, the physics of shear layer growth are expected to be similar in the two cases. 

By equating the timescale for shear-driven turbulence to bring hot gas into the mixing layer with the minimal cooling time of gas in the mixing layer, \citet{fielding_multiphase_2020} obtain an expression for the mixing layer thickness, $\delta$, (see their equation 4)
\be 
\label{eq:shear_width_D}
\frac{\delta}{\Rs} = \left(\frac{t_{\rm cool}\Vs}{\Rs}\right)^{3/2} \left(\frac{V_{\rm turb}}{\Vs}\right)^{3/2} \chi^{3/4}. 
\ee 
{\no}They find that $V_{\rm turb}\sim (0.1-0.2)\Vs$ independent of other parameters, such as the density contrast. A similar result was found for the turbulent velocities in mixing layers around cylindrical streams in the absence of radiative cooling \citep{M19}. In the context of the \citetalias{mandelker_lyalpha_2020} model described above, if we assume that \equ{shear_width_D} can be applied locally at every halocentric radius, this implies that $\delta/a\propto (x^{\beta}y/\mu)^{3/4}$, where $y=(\Vs(r)/\Vs(\Rv))^2$ is the stream velocity at radius $r$ normalized by its velocity at $\Rv$, squared. In practice, for $10^{12}\msun$ halos at $z\sim 1$, $\delta/a \sim (0.01-0.1)$ throughout the halo. For $\Rs\sim 0.25\Rv\sim 40\kpc$, this implies $\delta\sim 0.4-4\kpc$ near the outer halo, and slightly narrower towards the halo centre. This is comparable to our assumed values of $\delta\sim (1-2)\kpc$ for defining the CI interface gas in the CGM of our simulations, and can serve as a post-facto justification of this ad-hoc choice.

The simulations of \citet{ji_simulations_2019} have different resolution, initial perturbation spectrum, and cooling curve than those of \citet{fielding_multiphase_2020}. They also explore a different range of parameter space, and differ in their analysis methods. All these lead them to propose a different expression for the mixing layer thickness, based on their simulations. The main difference in their modeling is that they assume that pressure fluctuations induced by rapid cooling are what drive the turbulence in the mixing region, rather than the shear velocity. They suggest the following expression for the interface thickness (see their equation 27)
\be 
\label{eq:shear_width_J}
\delta \sim 750 \pc \left(\frac{\Lambda(T_{\rm mix})}{10^{-22.5}{\rm erg~s^{-1}}}\frac{n_{\rm s}}{3\times 10^{-4}\cmc}\frac{T_{\rm s}}{3\times 10^4\K} \right)^{-1/2},
\ee 
{\no}where we have normalized the cooling rate, density, and temperature by typical values found in our simulations (see \se{streams_sims}). In the context of the \citetalias{mandelker_lyalpha_2020} model described above, where the stream is isothermal with density and radius following \equs{rho_prof}-\equm{rs_prof}, this implies $\delta/a\propto \mu^{-1/2}$, which is nearly constant throughout the halo. This is comparable to our assumed interface thickness of $\sim (1-2)\kpc$ in the outer halo, but predicts a narrower interface layer closer to the halo centre, where the stream becomes narrower as well.

Importantly, even if the mixing layer thickness itself is unresolved, the mass entrainment rate and the associated stream deceleration and energy dissipation, are found to be converged at relatively low spatial resolution of $\sim 30$ cells per stream diameter, which is the scale of the largest turbulent eddies (\citetalias{mandelker_instability_2020}; see also \citealp{ji_simulations_2019,gronke_how_2020,fielding_multiphase_2020}). This is comparable to what is achieved in the VELA simulations. 

\begin{table*}
\begin{centering}
 \begin{tabular}{l c c c c c c c c c} 
 \hline\hline
VELA  & $T_{\rm hot}/\Tv$ & $\rho_{\rm s}/\rho_{\rm hot}$ & $n_{\rm s}~{\rm [10^{-4} \cmc]}$ & $R_{\rm s}/\Rv\:(n_{\rm s}=1)$ & $R_{\rm s}/\Rv\:(n_{\rm s}=3)$ & $V_{\rm s,tot}/\Vv$ & $V_{\rm s,rad}/V_{\rm s,tot}$ & $Z_{\rm s}/Z_{\rm \odot}$ & $Z_{\rm hot}/Z_{\rm \odot}$ \\
 \hline
 {\bf V07} & {\bf 2.7} & {\bf 85} & {\bf 3.2} & {\bf 0.42} & {\bf 0.24} & {\bf 1.02} & {\bf 0.15} & {\bf 0.08} & {\bf 0.58}   \\ 
 V08 & 1.38 & 22 & 2.7 & 0.49 & 0.28 & 0.82 & 0.52 & 0.06 & 0.32 \\
 V10 & 1.64 & 36 & 3.5 & 0.56 & 0.32 & 0.81 & 0.72 & 0.06 & 0.22 \\
 V21 & 2.04 & 20 & 1.6 & 0.35 & 0.20 & 1.01 & 0.50 & 0.06 & 0.34 \\
 V22 & 5.33 & 167 & 2.3 & 0.20 & 0.12 & 1.16 & 0.09 & 0.07 & 0.35 \\
 V29 & 1.73 & 20 & 1.8 & 0.45 & 0.26 & 1.00 & 0.59 & 0.06 & 0.26 \\
 \hline
 {\bf average} & {\bf 1.97} & {\bf 36} & {\bf 2.7} & {\bf 0.43} & {\bf 0.25} & {\bf 0.93} & {\bf 0.46} & {\bf 0.06} & {\bf 0.33} \\
 \hline\hline
 \end{tabular}
  \caption{Properties of the cold streams and the hot CGM at the halo virial radius, as inferred from our model, in the six VELA simulations examined in this work. From left to right we list the VELA index, the ratio of hot CGM temperature to the halo virial temperature, the density ratio between the cold stream and hot CGM, the volume density in the cold streams, the ratio of stream radius to halo virial radius assuming one stream in the halo, the average ratio of stream radius to halo virial radius assuming three streams, the ratio of stream velocity to the halo virial velocity, the ratio of stream radial velocity to total velocity, the metallicity in the cold streams, and the metallicity in the hot CGM. The range of these parameters found in the simulations is consistent with the predictions of \citetalias{mandelker_lyalpha_2020}.}
\label{tab:stream_props}
 \end{centering}
\end{table*}

\subsection{Comparison to Simulation Results}
\label{sec:streams_sims}

We now analyze the overall properties of the three identified states of gas in the VELA simulations. Each cell is assigned to one of the states (PI, CI-interface, or CI-bulk). PI gas is defined as in Section \ref{sec:definitions}. The `interface CI' gas cells are defined via the following two criteria (besides being CI). (1), they are within 2 kpc of a PI cell, as described in \ref{sec:3Ddistribution}, and (2), they have an OVI number density above 10$^{-13}$ cm$^{-3}$, to allow the interface to become smaller than 2kpc as the resolution improves in the inner halo. Any CI cell not classified as `interface CI' is classified as `bulk CI' instead. The first criteria is justified based on Equations \ref{eq:shear_width_D} and \ref{eq:shear_width_J}, and the second will be discussed in \se{other_ions}. 
In Figure \ref{fig:streams_profiles_1} we show the temperature, density and volume of each phase from 0.1 to 1.0$\Rv$. For each component, we fit a power-law relation to the profiles at $r>0.5\Rv$, and list the best-fit relation in the legends. We restrict ourselves to the outer half of the halo when fitting the profiles in order to minimize the effects of galactic feedback and of the non-radial orbit of the stream (see below), both of which are not accounted for in the analytic model of \citetalias{mandelker_lyalpha_2020} described in \se{streams_theory}. Indeed, in many cases, the profiles noticeably change around $r\sim (0.4-0.5)\Rv$, when these effects likely become important. 

In Figure \ref{fig:streams_profiles_1}, we see that the temperature of PI gas is $\gsim 3\times 10^4\K$ at $\Rv$, and decreases roughly as $r^{0.8}$ to a temperature of $\lsim 10^4\K$ at $0.1\Rv$. Nonetheless, at $r>0.4\Rv$, this gas is close to isothermal at $3\times 10^4\K$. The drop in temperature towards lower radii is due to increasing density (centre panel), shortening the cooling time and reducing the heating by the UV background. The bulk CI gas has a temperature of $\lsim 3\Tv$ at $\Rv$, increasing roughly as $T\propto r^{-0.5}$ towards $0.4\Rv\sim 60\kpc$. At smaller radii, the temperature increases sharply as hot outflowing gas from the galaxy becomes more prominent and the pressure rises (see \fig{GASPROPSslice}b and \fig{GASPROPSslice}d). This also corresponds to the radius where the OVI CI fraction sharply increases (\fig{ci-frac}). At $0.1\Rv$, the bulk CI gas reaches temperatures of $\sim 20\Tv$. These extremely large temperatures are likely dominated by hot feedback-induced outflows from the galaxy. The CI interface, which contains the vast majority of total CI gas mass (\tab{2}), has temperatures much closer to $\Tv$ throughout the CGM. The average temperature of the all CI gas is nearly isothermal at $\sim 2\Tv$. All in all, we find the temperature profiles of the PI gas and CI gas consistent with the expected behaviour for cold streams and the hot CGM, respectively, as described in \se{stream_halo}. 

The density in the PI gas near $\Rv$ is $\sim 3\times 10^{-4}\cmc$. This is consistent with the predicted densities of cold streams near $\Rv$ of $10^{12}\msun$ halos at $z\sim 1$, albeit towards the low-end of the expected range\footnote{The relatively low value for the density in this case results from the fact that in this particular galaxy, the hot CGM contains only $\sim 10$ percent of the baryonic mass within $\Rv$, rather than the fiducial value of $\sim 30$ percent assumed in \citetalias{mandelker_lyalpha_2020} (see their equation 24).}. 
The density increases towards the halo centre roughly as $r^{-2.3}$. This is much steeper than the density profile in the CI bulk, which scales as $r^{-1.5}$ outside of $\sim 0.4\Rv$, and has an even shallower slope at smaller radii. The steeper increase of the PI gas density towards the halo centre compared to the CI bulk, allows the cool phase to remain close to (albeit slightly below) pressure equilibrium throughout the halo, despite the decrease (increase) in the temperature of PI (CI bulk) gas towards the halo centre (see also \fig{GASPROPSslice}d). At $\Rv$, the PI gas is $\sim 85$ times denser than the CI bulk, consistent with the predicted density contrast between cold streams and the hot CGM \citepalias{mandelker_lyalpha_2020}. The CI interface also maintains approximate pressure equilibrium with the PI gas and the CI bulk throughout the halo, with density and temperature values roughly the geometric mean between those two phases. This is as expected for turbulent mixing zones (\equsnp{nmix}-\equmnp{Tmix}).

The volume occupied by the PI gas interior to radius $r$ scales as $r^{2.54}$, in agreement with \equ{Vol_s} given the slope of the CI bulk density profile. Assuming that the total volume of the PI gas is composed of $n$ streams, we can infer the typical stream radius by equating the right-hand-side of \equ{Vol_s} with $V_0/n$, where $V_0\sim 1.5\times 10^6 \kpc^3$ is the total volume of PI gas at $\Rv$ shown in \fig{streams_profiles_1}. The result is $\Rs/\Rv \sim 0.25$, $0.30$, and $0.40$ for $n=3$, $2$, and $1$ respectively. Most massive high-$z$ galaxies are predicted to be fed by 3 streams \citep{dekel_cold_2009,Danovich_2012}, with a single `dominant' stream containing most of the mass and volume. Visual inspection of VELA07 at $z=1$ reveals that $n=2$ is likely the best value (see \fig{sample}, and the top-right and bottom-right of \fig{GASPROPSslice}a). We also note that if the stream is not radial, but rather spirals around the central galaxy, as in \fig{sample}, the total stream volume will be larger than inferred from \equ{Vol_s}, and this can also be included by an effective $n>1$ for a single stream. Regardless, the inferred values of $\Rs/\Rv$ for $n=(1-3)$ are consistent with expectations \citepalias{mandelker_lyalpha_2020}. 

These results for the temperature, density, and volume of the three CGM phases lead us to conclude that we can associate the PI gas with cold streams, the CI bulk gas with a background hot halo, and the CI interface gas with a turbulent mixing layer forming between the two as a result of KHI \citepalias{mandelker_instability_2020}. While we have focused our discussion on VELA07, the other galaxies examined in this work exhibit very similar properties, and are all consistent with this association. We list their properties in \tab{stream_props}, all of which are consistent with the predictions of \citetalias{mandelker_lyalpha_2020}. To further solidify this point, we now examine the profiles of velocity and line-mass of the PI gas, and compare to predictions for the evolution of cold streams flowing through a hot CGM \citepalias{mandelker_lyalpha_2020}. 

The left-hand panel of \fig{streams_profiles_2} shows radial profiles of the total velocity magnitude for the three CGM components. The PI and CI interface gas both have velocities of $\sim \Vv$ at $\Rv$. While the velocity at $r>0.5\Rv$ is nearly constant, their velocity at $0.1\Rv$ is $\sim 1.6\Vv$, slightly less than the free fall velocity at this radius, which is $\sim 2.5\Vv$ assuming an NFW halo with a concentration parameter of $c\sim 10$. The CI bulk has velocities of order $\sim 2\Vv$ at $\Rv$, and increases by a similar factor between $\Rv$ and $0.1\Rv$. These super-virial velocities are due to strong winds (see \fig{GASPROPSslice}b), and are consistent with the super virial temperatures in this component seen in \fig{streams_profiles_1}.

In the middle panel of \fig{streams_profiles_2} we show profiles of the radial component of the velocity normalized to the total velocity at that radius, for the three CGM phases. The CI bulk gas is outflowing almost purely radially from $0.1\Rv$ to $\Rv$. The PI gas, on the other hand, is inflowing from $\Rv$ to $0.1\Rv$, but with a significant tangential component. This is consistent with models for angular momentum transport from the cosmic web to growing galactic disks via cold streams \citep{danovich_four_2015}. These tangential orbits can be inferred from \fig{sample}, where a stream can be seen spiralling in towards the central galaxy. Such orbits were not considered by \citetalias{mandelker_lyalpha_2020}, who only considered purely radial orbits for the streams. We therefore cannot strictly apply the predictions of their model to the stream dynamics within the halo. However, we expect that the model should work reasonably well in the region $r\gsim 0.5\Rv$, where the orbit is mostly along a straight line before the final inspiral begins. The magnitude of the radial component of the CI interface gas velocity is comparable to that of the PI gas. However, this component experiences both net inflow and outflow intermittently, likely depending on the orientation of the inflowing stream with respect to the outflowing bulk gas.

\begin{figure*}
    \includegraphics[trim={3.7cm 0.5cm 4.0cm 0cm}, clip, width=0.99\textwidth]{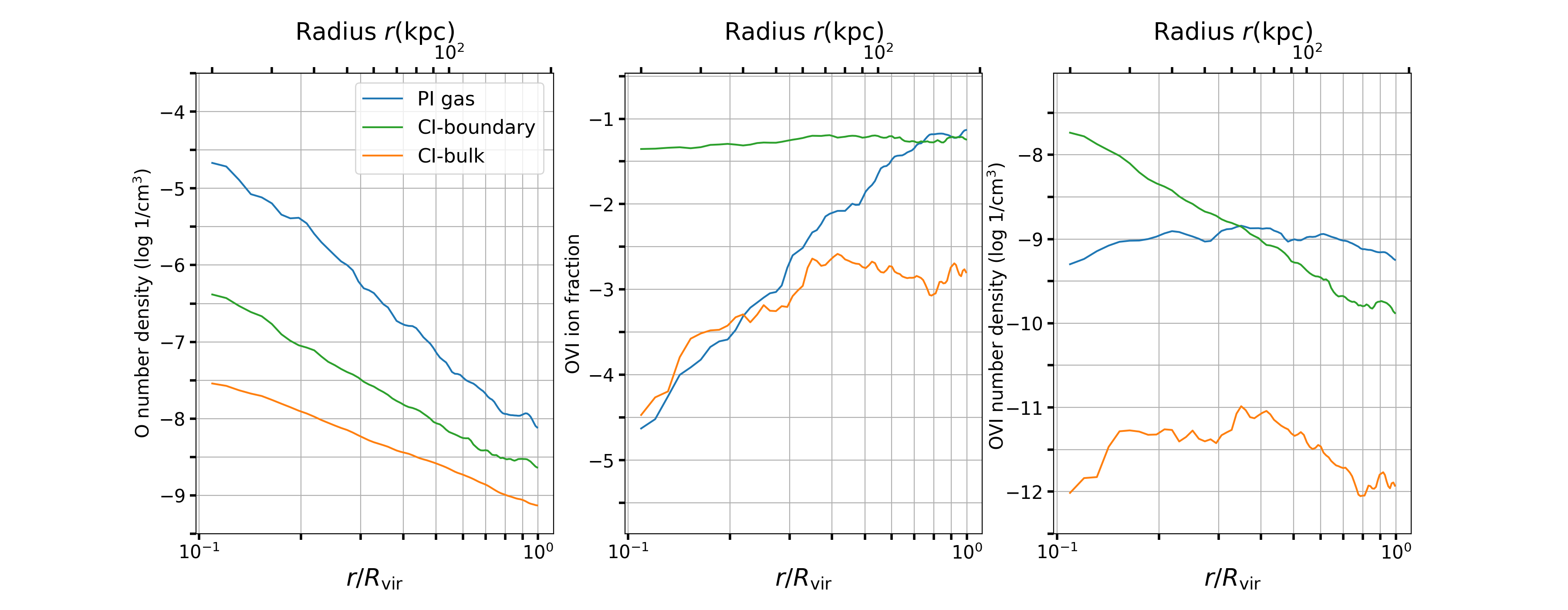}\par 
\caption{Radial profiles of OVI properties in the three OVI phases in the CGM of VELA07 at $z=1$. Line colours are as in \fig{streams_profiles_1}. \textit{Left:} Total oxygen number density, determined by total density, total metallicity, and overall oxygen abundance. \textit{Centre:} OVI ion fraction within each phase. 
\textit{Right:} OVI number density within each phase, which is in effect the product of the previous two panels. 
}
\label{fig:OVIstreamprops_profile}
\end{figure*}

In the right-hand panel of \fig{streams_profiles_2} we show the line-mass (mass-per-unit-length) of the three CGM components as a function of halocentric radius. The line-mass of the PI gas increases by $\sim (5-10)$ percent from $\Rv$ to $0.4\Rv$, comparable to the predictions from the model of \citetalias{mandelker_lyalpha_2020}. It then proceeds to increase rapidly, growing by more than a factor of 5 during the inspiral phase at $r<0.4\Rv$. We also note that at all radii, the line-mass of the CI interface gas is $\sim 5$ percent of the line-mass of the photo-ionized gas. This implies that the mass flux of hot gas being entrained in the stream is proportional to the stream mass, which is indeed predicted to be the case (\equnp{mass_rad_approx}). This strengthens our association of the PI gas and CI-interface gas with cold streams and the turbulent mixing layers that surround them, respectively.

\subsection{Suggested `Inflowing Streams' Model for OVI}
\label{sec:other_ions}

Since both substantial components of OVI (PI gas, and CI interface gas) are closely linked to the physical phenomenon of inflowing cold streams, as discussed above, we suggest that OVI absorption sightlines in the CGM, and possibly metal absorption spectra more broadly, should be modeled as a three-phase structure following Figure \ref{fig:model}. There are three phases to the CGM: Inside of the cool-inflow streams, their interface, and the outside bulk region. These streams, which narrow as they approach the galaxy, can be characterized geometrically as `spiraling cones', with a fit to their number $n$, their average cross-sectional radius $a(r)$, and their interface size $\delta(r)$. Internally, these streams would have a temperature, density, and metallicity which depends on $r$ as well. The properties of these streams will change with redshift, which could explain some of the differences between the $z\sim1$ data here and the lower-redshift COS-Halos results, including that the streams are expected to get wider as $z$ approaches 0 \citep{dekel_cold_2009}. 

\begin{figure*}
\begin{multicols}{2}
    \includegraphics[trim={0.7cm 0.8cm 0.7cm 0.05cm}, clip, width=0.99\linewidth]{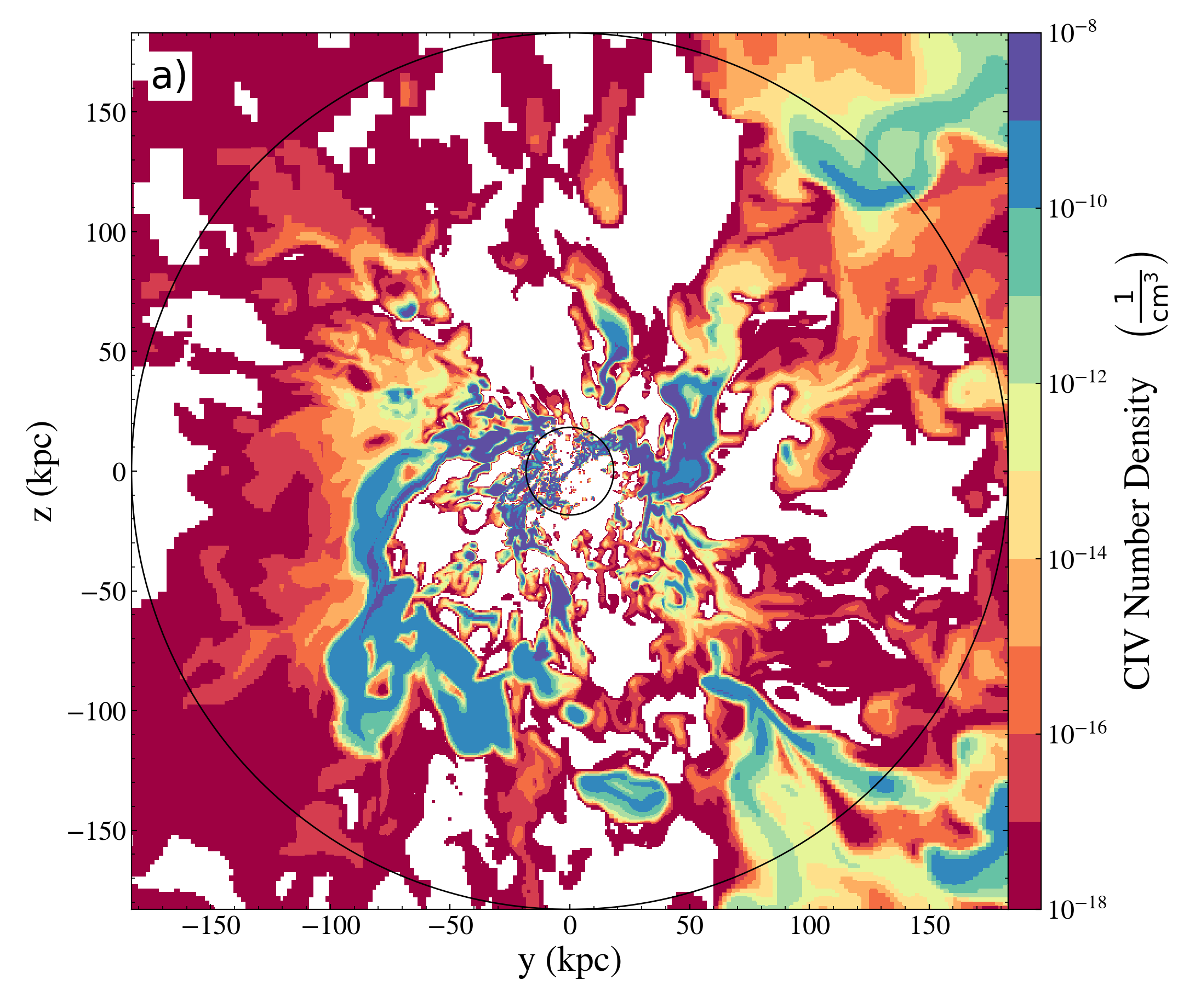}\par 
    \includegraphics[trim={0.7cm 0.8cm 0.7cm 0.05cm}, clip, width=0.99\linewidth]{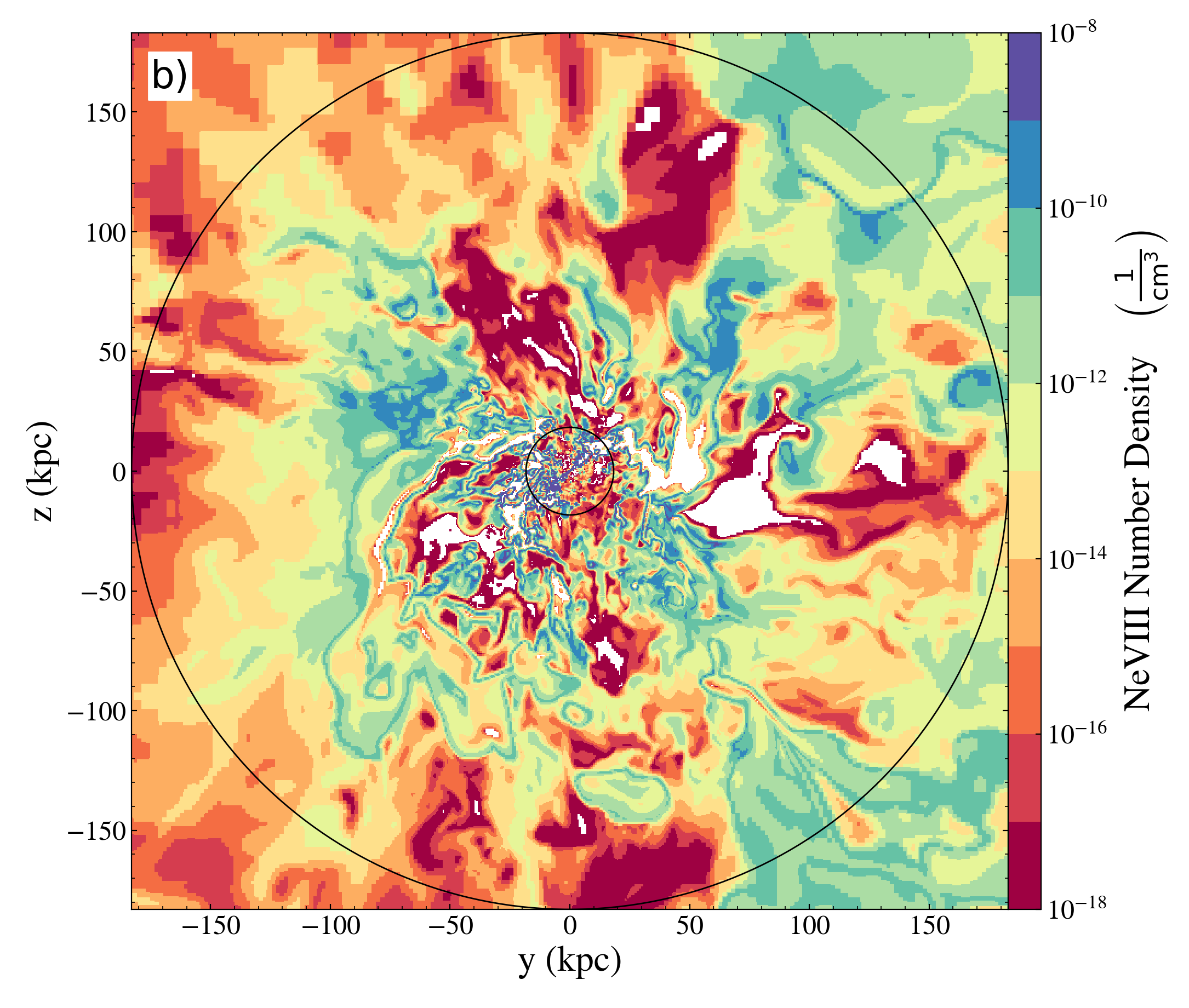}\par
\end{multicols}
\caption{Two different ions of lower and higher ionization than OVI, in the same slice as Figures \ref{fig:GASPROPSslice} and \ref{fig:OVIslice}. \textbf{(a):} CIV, \textbf{(b):} NeVIII. Note these images are of the same slice as Figure \ref{fig:OVIslice}, but with a slightly lower dynamic range, reflecting carbon and neon's lower abundances compared to oxygen. The same three phases, including the thin interfaces, are visible in these other ions.}
\label{fig:otherions}
\end{figure*}

Within each phase, we would suggest that each ion number density should fit to a power law with radius, with some exceptions as we will describe. An example of this is shown in Figure \ref{fig:OVIstreamprops_profile}. Here we see that the total oxygen within the streams, interface and bulk all increase as they approaches $r=0$, reflecting the increase in both density and metallicity there. In the streams, this increased density ends up lowering the OVI ion fraction so much that the total density of OVI within streams remains nearly constant throughout the CGM. At the same time, the interface layer gas maintains a constant ion fraction as its density increases, since in CI this fraction is nearly density-independent. Therefore, its OVI density increases to become higher than that of the PI gas in the inner halo. Finally, the bulk gas has both a low oxygen density and a low OVI fraction, and is irrelevant throughout the halo. Combining this plot with the volume plot in the right panel of Figure \ref{fig:streams_profiles_1} which showed that in the inner halo the interface tends to fill approximately the same volume as the stream it envelops, leads to the conclusions from this paper and \citetalias{roca-fabrega_cgm_2019} that PI gas is more significant in the outer halo, and that sightlines mostly intersect PI gas outside 0.3 $R_{\rm vir}$ (see \se{sightlines}). The second threshold (besides the requirement to be within 2 kpc of a PI cell) of $n_{\rm OVI}>$10$^{-13}$ cm$^{-3}$ for interface gas is shown here to be not too high of a threshold, as the OVI number density for CI bulk gas is actually generally still higher than 10$^{-13}$ cm$^{-3}$ and the interface is significantly higher, so its properties do not come primarily from selection bias. A higher threshold would decrease the cumulative volume in the interface, but otherwise would not significantly change its properties.

We briefly describe the procedure to fit any ion's 3D profile to this model. Since the temperature change with radius in each phase is significantly less than the density change (Figure \ref{fig:streams_profiles_1}), we would begin by assuming constant temperatures for the bulk, interface, and stream, with characteristic values as determined by \citet{begelman_turbulent_1990,gronke_growth_2018}; \citetalias{mandelker_instability_2020}; \citetalias{mandelker_lyalpha_2020}, and other alternatives. Given those temperatures, we determine using the procedure of Section \ref{sec:definitions} whether the ion will be PI, CI, or transitionary. If the ion is determined to be CI, its fraction (under the assumption of constant temperature) will be constant and its density will be therefore a constant times the phase density, unless this is lower than the critical CI density, in which case we will instead assume the CI contribution is a declining power law with decreasing radius. If the ion is determined to be PI, its ionization fraction $f_{X^i}$ (i.e. the fraction of the $i$th state of atom $X$) can be simplified by assuming a broken power law, with a positive power in $f-n$ space below some density, an approximately flat region, and then a negative power in $f-n$ space above some density. This decomposition is justified by examining the PI ions in Figure \ref{fig:ionfractions}, and images like it at other temperatures. If the breaks between these multiple power laws do not overlap with the density ranges within the three phases ($\approx 2$ dex, depending on which phase is under discussion), the ion number density itself will follow a power law: 
\begin{equation}
n_{X^i} \propto Z\cdot \rho \cdot f_{X^i},
\end{equation}
where $n_{X^i}$ is the number density of that ion state in that phase, $Z$ is the metallicity, and this equation will apply in any of the stream, interface, or bulk phases of the CGM. On the other hand, if the breaks between the power laws do overlap with the density ranges, the function will be much more complicated and probably cannot be well modeled. If the ion is at a `transitionary' temperature, it can be modeled as a broken power law with four segments, adding an additional flat curve at high density. This does not change the procedure, except to increase the likelihood that the model will break down due to the additional power law break. 

In this picture, OVI is unique only in that its line of distinction between PI and CI mechanisms, as we defined in Section \ref{sec:definitions}, happens to coincide with the temperature distinction between the streams and their interfaces. We have shown in Sections \ref{sec:streams_theory} and \ref{sec:streams_sims}, by comparing the phases defined by the OVI ionization mechanisms to theoretical studies of cold streams and their properties, that these inflowing streams are identifiable with the regions of PI OVI. There is no reason to believe that other commonly-observed ions should have a meaningful CI boundary layer on the edge of PI clouds, or indeed that they are PI within the cold streams, and CI outside of them. We show two other ions in Figure \ref{fig:otherions} as an example, one of which (CIV) has a lower ionization energy than OVI while the other (NeVIII) has a higher ionization energy. CIV appears here to be even more negligible outside the cold streams, and within the streams falls off more strongly with radius (see the top right and bottom right clouds within the slice). The interface layers have lower CIV density (green, instead of blue), as opposed to comparable or higher OVI density. 
On the other hand, NeVIII is not localized to the streams at all, but rather has a higher density in the bulk material, and is highlighted in the interface layer in particular, which has a higher NeVIII density than either the bulk or the stream. The fact that the same streams and interface layers identified in OVI are also visible in NeVIII, though with totally different relative ion densities, is further evidence that the stream interface layers are a real phenomenon in the simulation, even though they were detected using the definition of the OVI CI-PI cutoff and not their other physical properties. This dependence is summarized in the following list, which shows how this model can lead to vastly different distributions throughout the CGM for similar (e.g. lithium-like, or containing three electrons) ions. 

\begin{itemize}[leftmargin=*]
    \item For medium-ion states (e.g. CIV), we have $n_{\rm stream} \gg n_{\rm interface} \sim n_{\rm bulk}$. 
    \item For mid-to-high ion states (e.g. OVI), we have $n_{\rm stream} \sim n_{\rm interface} \gg n_{\rm bulk}$
    \item For high-ion states (e.g. NeVIII) we have $n_{\rm interface} \gtrsim n_{\rm bulk} \sim n_{\rm stream}$
\end{itemize}
We do not here include a prediction for low ion states. While these could be fit to this model, they are likely subject to resolution limits \citep{hummels_impact_2019}, so small clouds which are not produced in VELA could form a substantial contribution. The testable predictions of this model are that gas in mid-level ions reside in the inflows and can be detected all the way to the outer halo and beyond, while high ions (NeVIII) are significant throughout the bulk, and not strongly correlated to HI. 

\section{Summary and Conclusions}\label{sec:conclusion}

In this work we study properties of OVI in the CGM of 
$\sim 10^{12}\msun$ halos at $z\sim 1$ from 
the VELA simulations. We introduce a procedure for identifying all ions as photoionized or collisionally ionized, depending on the density, temperature, redshift, and assumed ionizing background, with negligible `overlap'. This causes low ions to convert from PI to CI at lower temperatures than high ions, resulting in large regions where some ions could be PI and others CI simultaneously. We run mock sightlines through the simulations and compare the results with data from observations, suggesting a toy model for use in future work. 

The main results of our analysis can be summarized as follows:
\begin{itemize}
\item{\bf Photoionized cool inflows:}  PI OVI is found entirely within filamentary cool inflows from outside the CGM. While they fill only a tiny fraction of the CGM volume, most of the OVI in the CGM is located inside them.
\item{\bf Collisionally Ionized Interface Layer:} The cool inflows have a warm-hot thin interface layer, which is the primary source of CI OVI. 
\item{\bf Low-density Collisionally Ionized Bulk:} The bulk of the CGM by volume is at a high enough temperature that OVI is CI, however this phase is negligible in terms of total OVI mass. This results in undetectably low CI-dominated column densities outside of the inner halo.
\item{\bf OVI sightlines are mostly PI in the outer halo of massive galaxies at $z\sim1$:} Since sightlines naturally probe the outer halo more than the inner, the cool inflows structure above leads to detectable OVI column densities being dominated by PI gas for all impact parameters outside $0.2-0.3R_{\rm vir}$.
\item{\bf Assumptions of spherical symmetry underestimate OVI median radius:} The non-spherical nature of the `cool inflows' model leads inverse Abel transformations (as in \citetalias{stern_does_2018}) to predict that the median OVI particle is located at around $0.6R_{\rm vir}$. However, this is an underestimate compared to the actual distribution of gas. Most OVI is therefore likely located very near, or beyond, the virial radius.
\item{\bf Inflows characterize the OVI structure of the CGM of massive galaxies at $z\sim1$:} We propose a model in which metal absorbers are characterized by their number densities in three distinct phases: inside cool inflows, outside the inflows in the bulk CGM volume, and in an interface layer between these two phases. This geometrical structure is characterized by the characteristic radius of the inflowing stream (which is itself a function of halocentric radius), $a(r)$, and the thickness of the interface layer, $\delta$. This model appears consistent with analytical predictions about the gas distribution from the interaction between cold streams and the hot CGM \citepalias{mandelker_lyalpha_2020}.
\item{\bf OVI is unique in tracing both the stream and interface:} While the three-phase cool streams structure we describe here is a general prediction for observations of the CGM, OVI has a PI-CI cutoff which matches the difference between the stream and interface conditions.
\end{itemize}

Future work will apply the same framework for distinguishing PI and CI gas to other ions. We are especially interested in CIV \citep{steidel_structure_2010,bordoloi_cos-dwarfs_2014} and NeVIII \citep{burchett_cos_2019,prochaska_cos_2019} surveys, at redshifts $z\gtrsim1$, so their actual column density values could be compared to those in VELA. This would let us further develop this three-density CGM model with $n_{\rm stream}$, $n_{\rm interface}$, and $n_{\rm bulk}$. We will also follow the same idea in other simulations that reach lower redshifts and different mass ranges, but which have good enough resolution to show these cool inflows in the CGM. Finally, a comparison with the new generation of the same VELA galaxies (Ceverino et al. in prep.) will allow us to directly compare the effects of increased feedback on the CGM with the same initial conditions. 
\section*{Acknowledgements}
Partial support for CS was provided by grant HST-AR-14578 to JP from the STScI under NASA contract NAS5-26555. SRF acknowledges support from a Spanish postdoctoral fellowship, under grant number 2017- T2/TIC-5592. SRF also acknowledges financial support from the Spanish Ministry of Economy and Competitiveness (MINECO) under grant number AYA2016-75808-R, AYA2017-90589-REDT and S2018/NMT-429, and from the CAM-UCM under grant number PR65/19-22462. NM acknowledges support from the Klauss Tschira Foundation through the HITS Yale Program in Astrophysics (HYPA), and from the Gordon and Betty Moore Foundation through Grant GBMF7392. JS is supported by the CIERA Postdoctoral Fellowship Program. The VELA simulations were performed at the National Energy Research Scientific Computing Center (NERSC) at Lawrence Berkeley National Laboratory, and at NASA Advanced Supercomputing (NAS) at NASA Ames Research Center. DC is a Ramon-Cajal Researcher and is supported by the Ministerio de Ciencia, Innovaci\'{o}n y Universidades (MICIU/FEDER) under research grant PGC2018-094975-C21. Partial support for AD comes from DIP STE1869/2-1 GE625/17-1 and ISF 861/20. BW and RD carried out this research under the auspices of the Science Internship Program (SIP) at the University of California Santa Cruz, which also provided partial funding to CS. Technical help was provided by Cameron Hummels, Britton Smith, Nathan Goldbaum, Matt Turk, and the rest of the \textsc{yt} community.  We also benefitted from helpful discussions with Joe Burchett, Sandra Faber, Yakov Faerman, Max Gronke, David Koo, Daisuke Nagai, S. Peng Oh, J. Xavier Prochaska, Rahul Sharma, and Sally Zhu.  Finally, we thank the comments of an anonymous referee which improved several key points of the paper.
\section*{Data Availability}
Data underlying this article is available at \url{https://github.com/claytonstrawn/quasarscan}.



\bibliographystyle{mnras}
\bibliography{strawn20bib.bib}




\bsp	
\label{lastpage}

\end{document}